\documentclass[a4paper,12pt,reqno]{amsart}

\usepackage{a4}
\usepackage{verbatim}
\usepackage{comment}
\usepackage{color}
\usepackage{todonotes}
\usepackage{doi}

\usepackage{amssymb,stmaryrd}
\usepackage{color}
\usepackage{graphicx}
\usepackage{floatflt}
\usepackage{float} 		%damit H funktioniert bei bildern
\usepackage{amsthm}
\usepackage{amscd}
\usepackage{hyperref}

\newtheorem{definition}{Definition}

\newtheorem*{main}{Main Result}
\newtheorem*{theorem*}{Theorem}

\newtheorem*{lemma*}{Lemma}
\newtheorem{remark}[definition]{Remark}

\newtheorem*{corollary*}{Corollary}

\def\XXint#1#2#3{{\setbox0=\hbox{$#1{#2#3}{\int}$}
		\vcenter{\hbox{$#2#3$}}\kern-.5\wd0}}

\makeatletter
\@addtoreset{definition}{section}
\@addtoreset{equation}{section}
\makeatother

\DeclareMathOperator{\Res}{Res}

\allowdisplaybreaks[4]

\title[GW/DT invariants and 5D BPS indices for strips from TR]{GW/DT invariants and 5D BPS indices for strips from topological recursion}

\author{Sibasish Banerjee}

\address{IHES, 35 Route de Chartres, 91440 Bures-sur-Yvette, France}

\email{sbanerjee@ihes.fr}

\author{Alexander Hock}

\address{Section of Mathematics, University of Geneva, Rue du Conseil-G\'{e}n\'{e}ral 7-9,
	1205 Geneva, Switzerland} 
\email{alexander.hock@unige.ch}

\author{Olivier Marchal}

\address{Universit\'{e} Jean Monnet, Institut Camille Jordan UMR 5208, Institut Universitaire de France, Les Forges 2, 20 Rue du Dr Annino,
42000 Saint-Etienne, France}

\email{olivier.marchal@univ-st-etienne.fr}

\begin{document}
\maketitle

\begin{abstract}
    Topological string theory partition function gives rise to Gromov-Witten invariants, Donaldson-Thomas invariants and 5D BPS indices. Using the remodeling conjecture, which connects Topological Recursion with topological string theory for toric Calabi–Yau threefolds, we study a more direct connection for the subclass of strip geometries. In doing so, new developments in the theory of topological recursion are applied as its extension to Logarithmic Topological Recursion (Log-TR) and the universal $x$--$y$ duality. Through these techniques, our main result in this paper is a direct derivation of all free energies from topological recursion for general strip geometries. In analyzing the expression of free energy, we shed some light on the meaning and the influence of the $x$--$y$ duality in topological string theory and its interconnection to GW and DT invariants as well as the 5D BPS index.
\end{abstract}

\section{Introduction and summary}
\label{sec:intro}

In this work, our main focus lies on understanding the connection among various kinds of invariants, motivated by enumerative geometry and physics,  associated with Calabi-Yau threefolds (CY). Namely, the Gromov-Witten (GW) invariants arising as enumerative invariants in the context of counting stable maps from a Riemann surface $\Sigma_{g}$ with genus $g$ into the CY ($\mathcal{Y}$), in the class $\beta \in H_2(\mathcal{Y},\mathbb{Z})$, on the one hand. Through the ``remodeling conjecture'' \cite{Bouchard:2007ys}, (several aspects of which have been firmly established as mathematically proven already \cite{Eynard:2012nj,Fang:2013dna,Fang:2016svw}) the  genus $0$ GW invariants can be associated with the B-model geometry of the mirror $\mathcal{X}$, as periods/variation of Hodge-structure, whereas, the higher genera invariants are determined by topological recursion, the main technique that features in this paper. On the other hand, we consider the Donaldson-Thomas (DT) invariants  for a CY threefold $\mathcal{Y}$, associated with the counting of coherent sheaves, with stability determined by polarization \cite{MR2951762,MR2600874}. A relation between GW invariants and DT invariants is already known due to \cite{Maulik:2003rzb,Maulik:2004txy}. As we will see, DT invariants are also closely related to the counting of certain BPS indices in five-dimensional supersymmetric gauge theories, engineered by  CY threefolds \cite{Nekrasov:2014nea,Nekrasov:2016qym,Banerjee:2018syt}.  A relation between BPS indices and topological recursion was already studied for other geometries in \cite{Iwaki:2020efz,Iwaki:2021zif}. In the following, we will denote as $\mathcal{X}$ the mirror of the CY threefold $\mathcal{Y}$. In fact, the same counting problem through mirror symmetry can be formulated as the counting of stable special Lagrangian A-branes in the Calabi-Yau threefold ($\mathcal{X}$). 

The DT invariants correspond to the counting of BPS states in the 5D $\mathcal{N}=1$ theory obtained by compactifying M-theory on a CY threefold. Starting from M-theory on $\mathcal{Y}\times S^1\times\mathbb{R}^4$, one arrives at the 5D theory, defined by the geometry of the CY as $T^{\textrm{5D}}[\mathcal{Y}]$ on ${\mathbb{R}}^4\times S^1$. Now consider M5-branes wrapping  $\mathcal{C}_4\times S^1\times \mathbb{R}$ and M2-branes wrapping $\mathcal{C}_2\times {\mathrm{pt}}\times \mathbb{R}$, where $\mathcal{C}_i$ are holomorphic submanifolds in $\mathcal{Y}$ ($i$ corresponds to real dimension) and the second component and the third component indicate the submanifolds that they wrap inside $S^1$ and $\mathbb{R}^4$. Reducing on $S^1$ one obtains a bound state of D4-D2-D0 branes in type IIA string theory corresponding to an object of the derived (bounded) category of coherent sheaves $D^bCoh(\mathcal{Y})$ \cite{Dijkgraaf:2006um,Kontsevich:2000yf,kontsevich1995homological,Nekrasov:2014nea}. At the same time in $T^{\textrm{5D}}[\mathcal{Y}]$, the corresponding BPS objects are ``monopole strings'' wrapping $S^1\times \mathbb{R}$ and ``instanton particles'' wrapping $\mathbb{R}$. This relates the computation of the BPS index for this theory, to the computation of the ``rank zero DT invariants'' for the Calabi-Yau threefold $\mathcal{Y}$. In this work, for a specific class of geometries, we will provide these 5D BPS indices $\Omega$, starting from the GW partition function.

The type IIA string theory setup for the count of the $\Omega$ is related to the type IIA setup of the Gopakumar-Vafa (GV) through ``9-11'' flip \cite{Dijkgraaf:2006um}. Starting from the background $\mathcal{Y}\times S^1\times \mathbb{R}^4$, one deforms the $\mathbb{R}^4$ as a Taub-Nut as $\mathbb{R}^3\times S^1_{\mathrm{TN}}$. Reducing on the circle $S^1_{\mathrm{TN}}$ now, one obtains the relevant type IIA frame where D2 and D0 branes lift to M2 branes and Kaluza-Klein momenta in M-theory.  It is possible to include D4 branes which lift to M5 branes. However, for the class of CY threefolds $\mathcal{Y}$ that we will consider, there will be no compact 4-cycle, so we omit their description here. 

The GV invariants are intimately related to the GW invariants. This is well-known from both physics and mathematics perspectives \cite{MR1363062,pandharipande2003questionsgromovwittentheory}. The GW invariants however appear naturally in the context of A-model where one counts stable maps $\phi : \Sigma_g \to \mathcal{Y}$. Then the path integral in the A-model localizes to the holomorphic maps which correspond to closed algebraic curves in $\mathcal{Y}$. We will return to their geometric meaning in a short while, after describing the class of the CY threefolds in which our interest lies. 

We consider toric CY threefolds $\mathcal{Y}$ which in the A-model side appear as a symplectic quotient $\mathcal{Y} = \mathbb{C}^{k+3}// U(1)^k$. The moment map equations give rise to a toric diagram which defines $\mathcal{Y}$. The mirror B-model description was provided in \cite{Hori:2000kt}, which was $\mathcal{X} = \{A(e^x,e^y) = uv\} \subset \mathbb{C}^4$. The mirror Calabi-Yau $\mathcal{X}$ then can be considered as a conic fibration on a Riemann surface defined through $A(e^x,e^y) = 0$. In the particular class of examples that we consider, we focus on the ``strip geometries''  which belong to the class of ``toric trees'' \cite{Forbes:2005xt,Panfil:2018faz,Iqbal:2004ne}. They do not have any compact 4-cycle and the ``mirror curve'' is given as follows \cite{Panfil:2018faz}
\begin{align}\label{spectralcurveintro}
    A(e^x,e^y)=(1-e^y)\prod_{j=1}^s(1-\beta_j e^y)+(-1)^f e^xe^{y(1+f)} \prod_{j=1}^r(1-\alpha_j e^y)=0,
\end{align}
where  in $\alpha_i,\, i=1,...,r$ and $\beta_j,\, j=1,...,s$ are certain combinations of complex structure parameters $Q_i$ defined through equation \eqref{alphaibetaj} and are pairwise different throughout this article.
%{\color{violet}These parameters are defined as certain combinations of the complex structure parameters $Q_i$ of the mirror curve through equation \eqref{alphaibetaj}}. 

One way to relate the problem of counting 5D indices $\Omega$ (which are related to the DT invariants) is making the problem slightly more complicated through introducing a toric brane, which we will refer to as Aganagic-Vafa (AV) brane \cite{Aganagic:2000gs,Aganagic:2001nx,Bouchard:2007ys}. Consider an M5 brane wrapping $L\times S^1\times \mathbb{R}^2$, where $L$ corresponds to a special Lagrangian A-brane with topology $\mathbb{C}\times S^1$. It was argued in \cite{Aganagic:2000gs} that the complexified moduli space of this toric brane is the mirror curve corresponding to the mirror CY threefold $\mathcal{X}$.
This is at the heart of the remodeling conjecture proposed in \cite{Bouchard:2007ys} connecting the B-model computation to topological recursion. Our computations in this paper clarify certain aspects of it, in the context of strip geometries. 

However the presence of the AV brane gives rise to a new BPS sector corresponding to a 3D $\mathcal{N}=2$ theory engineered by the codimension two defect coupled to the 5D theory, giving rise to the 3D-5D BPS sector. These were partly understood in \cite{Banerjee:2018syt,Banerjee:2019apt,Banerjee:2020moh}. Generalizing the works of \cite{Gaiotto:2012rg}, it was shown how to compute the DT invariants of $\mathcal{X}$ starting from $(\mathcal{Y},L_{\mathrm{AV}})$ through the procedure of ``nonabelianization''. This later led to a geometric definition of DT invariants corresponding to objects in the Fukaya category $D^bFuk(\mathcal{X})$ which are compactly supported special Lagrangian A-branes, \cite{Banerjee:2022oed}, providing evidence for a proposal of Joyce \cite{MR1941627}. An alternative perspective emerged in \cite{Jeong:2025yys} and more will be clarified in an upcoming work \cite{Banerjee:2026mwg}, whereas the computations here will be extended for the case of open GW invariants and connections to open DT invariants will be clarified in \cite{Banerjee:2025shz}.  

As mentioned above,  we focus on toric tree Calabi-Yau (CY) threefolds that do not have compact 4-cycles given by a set of $\mathbb{P}^1$'s which have resolutions as either $\mathcal{O}\oplus \mathcal{O}(-2)\to \mathbb{P}^1$ or $\mathcal{O}(-1)\oplus \mathcal{O}(-1)\to \mathbb{P}^1$. The topological string partition functions for such geometries are known \cite{Iqbal:2004ne,Ooguri:2010yk}. Let us recall them here: 
\begin{equation}\label{genZtop}
Z_{\mathrm{top}} = \prod_{l=1}^\infty (1-q^l)^{-\frac{l\cdot \chi}{2}} \prod_{1\leq i_1<i_2<\cdots<i_n\leq \chi-1} \big(1-Q_{i_1}Q_{i_2}\cdots Q_{i_n}q^l)^{-s_{i_1}s_{i_2} \cdots s_{i_n} l},
\end{equation}
where we put $s_{i_k}=-1$ if the $\mathbb{P}^1$ was resolved by $\mathcal{O}(-1,-1)$ and $s_{i_k} = +1$ if the resolution is $\mathcal{O}(0,-2)$. Here $\chi$ is the Euler characteristic of the CY and $Q_i$ are the complex structure parameters. Looking at the partition function itself, we can write down the toric tree unambiguously, because the topological string partition function can be built up just from the gluing data in this case. Expanding the partition function in terms of the free energy by $Z=\exp \sum_g\hbar ^{2g-2}F_g$ with $q=e^\hbar$ yields for $g\geq 2$
%In this notation one of the main results of this paper, the genus $g\geq 2$ free energy from \eqref{freeenergyresult}, will be given as 
\begin{equation}\label{freeenergyintro}
    F_g=-\frac{\chi}{2} \frac{B_{2g}B_{2g-2}}{ 2g (2-2g)\cdot (2g-2)!}-\sum_{1\leq i_1<i_2 \cdots<i_n\leq \chi-1 } s_{i_1}\cdots s_{i_n} \frac{B_{2g}\mathrm{Li}_{3-2g}(Q_{i_1} \cdots Q_{i_n})}{2g(2g-2)!},
\end{equation}
 where $\textrm{Li}_s (z) = \sum_{m=1}^\infty \frac{z^m}{m^s}$ for $\mathrm{Re}(s)>1$, and defined on the whole complex plane by analytic continuation. The main result is the derivation of the free energy \eqref{freeenergyintro} from the spectral curve \eqref{spectralcurveintro} directly via topological recursion.

In the curve \eqref{spectralcurveintro} we have
\begin{equation}\label{alphaibetaj}
  Q_{i_1} \cdots Q_{i_n} =
    \begin{cases}
      \alpha_k & \text{if} \,\, s_{i_1} \cdots s_{i_n}=-1\\
      \beta_k & \text{if} \,\, s_{i_1}\cdots s_{i_n}=+1,
    \end{cases}       
\end{equation}
for some $k \in \{1,...,\chi-1\}$ depending on the structure of the strip geometry. The 5D BPS invariants depend on the charge $\Gamma(\mathcal{E}) = {\textrm{ch}}(\mathcal{E})\sqrt{\textrm{Td}(\mathcal{X})}$, for some coherent sheaf $\mathcal{E}$, $\mathcal{X}$ being the CY threefold in the B-model side. From this one can read off the charge vectors $\gamma$ which define the D4-D2-D0 brane-content. Using our proposal, we read off $\Omega(\gamma)$ from \eqref{freeenergyintro} (in these examples D4-brane charge is zero).

 As an example, consider the five punctured sphere obtained from gluing two copies of $\mathcal{O}(-1)\oplus \mathcal{O}(-1) \to \mathbb{P}^1$, with $\alpha_{1,2} \neq 0$. This is called suspended pinched point (SPP) in \cite{Ooguri:2010yk} and \cite{Mozgovoy:2020has}. The curve from \cite{Ooguri:2010yk} reads as 
\begin{equation}
\mu e^{2y}+ e^{x+y} + e^x + (1+Q\mu) e^y + Q = 0
\end{equation}
which matches \eqref{spectralcurveintro} after transforming $e^x \to - e^{-x}/Q$ and $e^y \to e^{-y}$ with $\alpha_1 = 1/Q, \alpha_2 = \mu$, for $f=0$. The GW partition function defined as $F_{\geq 2}=\sum_{g\geq 2} \hbar^{2g-2} F_g$ has zero radius of convergence in $\hbar$. So, we would need to perform the analysis of \cite{Alim:2021mhp} to define various Stokes sectors. In one of those sectors one can recover the partition function of \cite{Ooguri:2010yk}. 
We have
\begin{equation}\label{FgSPP}
    F_g^{\textrm{SPP}}=- \frac{3 B_{2g}B_{2g-2}}{ 4g (2-2g)\cdot (2g-2)!}-\frac{B_{2g}\mathrm{Li}_{3-2g}(Q\mu)}{2g (2g-2)!}
    +\frac{B_{2g}\mathrm{Li}_{3-2g}(Q)}{2g (2g-2)!}+\frac{B_{2g}\mathrm{Li}_{3-2g}(\mu)}{2g (2g-2)!},
\end{equation}
from which we read off that there are three kinds of stable D2-branes in the type IIA side, with classical (not instanton-corrected) periods $\log Q$, $\log \mu$ and $\log (Q\mu)$ respectively, which can be  computed from the associated Picard-Fuchs equations as well \cite{Forbes:2005xt}. The GV invariants then can be read off as 
\begin{equation}
n^{(0)}_\mu = n^{(0)}_Q = +1, \quad n^{(0)}_{Q\mu} = -1
\end{equation}
and all others are zero, where we denote $n^{(g)}_\beta$ as the GV invariants, $g$ being the genus and $\beta$ the corresponding curve class in $\mathcal{X}$. It was shown in \cite{Banerjee:2019apt}, that exponential networks associated to such geometries compute the GV invariants as $\Omega(\beta + kD0) = n^{(0)}_\beta$.  In \textsection \ref{sec:DT}, we use the following notation for the 5D indices which for these cases can be read off from the GV invariants, as 
\begin{equation}\label{omegaintro}
    \begin{cases}
      \Omega (nD0) = -\chi_{\mathrm{SPP}} = -3, n\geq 1 ;\\
      \Omega(D2_{Q_{i_1 \cdots i_n}}-kD0) = -s_{i_1} \cdots s_{i_n}, \, k\geq 0; \\ 
      \Omega(\overline{D2}_{Q_{i_1\cdots i_n}}-kD0) = -s_{i_1} \cdots s_{i_n}, \, k\geq 0;
    \end{cases}       
\end{equation}
where we label the D2 brane charges with the corresponding classical periods and the hyphen denotes the bound states of those D2 branes with several D0's. Let us also report that the topological string partition function from \cite{Ooguri:2010yk} whose asymptotic expansion in $\hbar$ is \eqref{FgSPP} 
\begin{equation}
Z^{\mathrm{SPP}}_{\mathrm{top}} \, 
= \prod_{k=1}^\infty \frac{(1-Qq^k)^k \, (1-\mu q^k)^k}{(1-q^k)^{\frac32 k}(1-Q\mu q^k)^k}.
\end{equation}
This clearly reproduces \eqref{omegaintro} in the ``DT chamber''. \\

In this work, our main result is the following. We use topological recursion which is a universal recursion procedure \cite{Eynard:2007kz} to compute an infinite family of multidifferentials $\omega_{g,n}$ on $n$ copies of the spectral curve, which actually have their own intrinsic definition that applies to more general classes of curves compared to the mirror curves. We will derive the free energy \eqref{freeenergydefinition} exactly for the geometries in question (``strip geometries'') starting from \eqref{spectralcurveintro}. This is strictly a B-model computation. However, by the remodeling conjecture \cite{Bouchard:2007ys}, this is related to the generating function of the closed GW invariants in the A-model, for the CY threefold $\mathcal{Y}$. 

Starting from the GW partition function \eqref{freeenergyresult}, we set $q=e^{\hbar}$ to obtain it in the Gopakumar-Vafa form, as was also observed in \cite{MR2222528}. Following the computation in \cite[Appendix A]{Banerjee:2019apt}, one obtains the DT partition function from \eqref{DTpartitionfn} as 
\begin{equation}\label{DTdef}
Z\vert_{[Q_i]\geq 0}
= \sum_{n\geq 0,k_i\geq 0} DT_{D6-kD2_i-nD0} Q_i^{k_i} q^n.
\end{equation}

Having said this, let us briefly mention the efficacy of topological recursion for the computation of the DT invariants. We will provide more details in the following section, concretely for the geometries at hand. As stated above, topological recursion (TR) \cite{Eynard:2007kz} provides a universal framework to associate an infinite family of multidifferentials $\omega_{g,n}$ indexed by $g,n\in \mathbb{Z}_{>0}$ to certain initial data, the spectral curve~\eqref{spectralcurveintro}. The actual computation of TR is recursive in $2g+n-2$ in the sense that $\omega_{g,n}$ is obtained from all $\omega_{g',n'}$ with $2g'+n'-2<2g+n-2$. We are mainly interested here in the free energy denoted by $\omega_{g,0} = F_g$ corresponding to the closed string amplitudes. Since TR has this recursive nature, it is almost impossible to obtain closed exact expressions for $\omega_{g,n}$. However, recent developments in TR regarding the so-called $x$--$y$ duality \cite{Alexandrov:2022ydc,Hock:2022wer} provide a new tool for actual computations avoiding the recursive procedure. We will apply this new machinery to derive the closed formulas for the free energy of the mirror curve~\eqref{spectralcurveintro}. This can be implemented here efficiently, because the $x$--$y$ dual multidifferentials $\omega_{g,n}^\vee$ are given by an explicit formula due to $y$ being unramified. In doing so, we have to apply an extension of TR which is called logarithmic topological recursion (Log-TR) \cite{Alexandrov:2023tgl,Hock:2023dno}. Thus, we demonstrate that for strip geometries, using TR one can derive the GW partition functions from which one can recover the corresponding DT partiton functions, matching with for example \cite{Ooguri:2010yk}. This sheds light on the interpretation of the $x$--$y$ duality in topological recursion in the context of topological string theory and a connection to DT invariants and 5D BPS indices, at least for strip geometries.

As a remark, we would like to point out that by the remodeling conjecture \cite{Bouchard:2007ys}, $\omega_{g,n}$'s encode the maps from a Riemann surface with genus $g$ and $n$ marked points to $\mathcal{X}$ as $\phi:(\Sigma_{g,n},\partial\Sigma_{g,n}) \to (\mathcal{X},L)$. We will come to their computations in a work in preparation \cite{Banerjee:2025shz}, for the case of strip geometries. We will be able to explicitly reproduce the open  DT-partition functions \cite{Panfil:2018faz}, and we will also propose a quantization program using (Log-)TR and $x$--$y$ duality via \cite{Hock:2025sxq}. 

Finally, we would like to mention the class of CY geometries analyzed in this paper. The strip geometries have been studied using several methods before from different perspectives. In \cite{Banerjee:2019apt} for example, exponential networks were invoked, whereas in \cite{Mozgovoy:2020has}, the computations were done using the 5D BPS quivers which were derived from the B-model side. They were studied using matrix models in \cite{Ooguri:2010yk}, or using techniques from crystal melting in \cite{Sulkowski:2009rw}. The closed topological string partition function was also derived in \cite{Iqbal:2004ne} by gluing topological vertices. In fact their connection to Nekrasov's functions \cite{Nekrasov:2003af} were found in \cite{Iqbal:2004ne} and was checked as well in \cite{Eguchi:2003sj}. Therefore, there are several results with which we can match our main ingredient that we computed in this paper, the genus $g$ free energy \eqref{freeenergyintro}.\\

{\it The novelty of this work is that it  uses the “remodeling conjecture” of \cite{Bouchard:2007ys} to derive explicit free energy expressions for strip geometries. The key technical tool is the application of $x$–$y$ duality, which constitutes the main methodological ingredient of this paper. This approach allows us to access the generating functions of Gopakumar–Vafa invariants in a direct way, and thereby compute the associated 5D BPS indices. By comparing with established A-model computations—such as those obtained from 5D BPS quivers or crystal melting models—our B-model derivation provides a concrete test of mirror symmetry. In addition, this method offers an alternative computational framework to exponential networks, and enables nontrivial consistency checks of the results obtained there.}\\

In \textsection \ref{sec:TR}, we provide details of the procedure of implementing TR for strip geometries. In \textsection \ref{sec:DT}, we will provide some concrete examples and comment on the physical meaning of the coefficients of the free energy. Thus, comparing literally \eqref{freeenergyintro} with the result \eqref{freeenergyresult} obtained from topological recursion. In Appendix \ref{appendixa}, we relegate some delicate details about residues of polylogarithms appearing in the TR computation.

\subsection*{Further directions}
There are several further directions one can pursue, building upon the observations and computations of this article. 

First, one may consider mirror curves from other classes of toric trees as $\mathbb{C}^3/(\mathbb{Z}_m\times \mathbb{Z}_n)$~\cite{Ooguri:2010yk}. Those mirror curves will have in TR a nontrivial dual side, that is, all $\omega_{g,n}^\vee \neq 0$, but the free energy and the partition function still admit a nice expression~\cite{Ooguri:2010yk}. Furthermore, one can consider mirror curves of higher genus, thus having nontrivial 4-cycles. To derive the partition function or the free energy, one has to use the so-called non-perturbative topological recursion, which, however, is not yet properly developed in the context of Log-TR. 

Second, one can study the quantum curve arising from TR for strip geometries, which will appear in our upcoming work ~\cite{Banerjee:2025shz}. One should also investigate further geometries with toric diagrams that are not strips. This will provide a direct connection between TR and open DT invariants, following~\cite{Panfil:2018faz}. In doing so, it might be possible to understand a correspondence between TR and quivers. 

A third important aspect in topological string theory, and also in Chern--Simons theory, is the idea of gluing, for instance in the sense of gluing toric diagrams. For strip geometries, gluing just amounts to adding $x \to x \pm \log (1-\beta z)$. This transformation, by the latter equation~\eqref{omgdual}, induces a very simple change in the dual $\omega_{g,n}^\vee$ in TR. From those correlators, the procedure of gluing can be derived for $\omega_{g,n}$ via $x$--$y$ duality, which will be completely nontrivial. This may give rise to a general understanding of gluing in TR. 

Last but not least, refined GW invariants are well known but not yet obtainable from TR for toric Calabi--Yau threefolds. By formula~\eqref{fgcompute}, it is actually possible to derive the refined GW invariants for strip geometries from this formula by adding formally $\epsilon_1^{2g_1-1}$ to $\omega_{g_1,1}^\vee$ and $\epsilon_2^{2g_2-1}$ to $\omega_{g_2,1}$. The unrefined case is $\hbar =\epsilon_1=-\epsilon_2$. Following exactly the computation in Appendix \ref{appendixa}, we recover refined GW invariants \cite{Iqbal:2007ii}. However, how to modify the original definition of TR to achieve this directly from $\omega_{g,n}$ is not clear to us. It is even not clear if the $x$--$y$ duality should hold in a refined setting. This approach to refinement should clearly be distinguished from the one in~\cite{Kidwai:2022kxx}.

\subsection*{Acknowledgment}
Special thanks go to Kohei Iwaki who first brought this problem to our attention. We also would like to thank Nafiz Ishtiaque and Saebyeok Jeong for discussions about related matters. We thank the organizers of the trimester program``Higher Rank Geometric Structures" at the Institut Henri Poincaré, Paris, where the collaboration ensued. SB acknowledges the working conditions and hospitality provided by CERN and by the University of Geneva. The work of S.B.
has been supported in part by the ERC-SyG project “Recursive and Exact New Quantum
Theory” (ReNewQuantum), which received funding from the European Research Council
(ERC) under the European Union’s Horizon 2020 research and innovation program, grant
agreement No. 810573.  A.H. was funded by
	the Deutsche Forschungsgemeinschaft (DFG, German Research
	Foundation) -- project ID 551478549 and by the Swiss National Science Foundation (SNSF) through the
Ambizione project “TRuality: Topological Recursion, Duality and Applications”
under the grant agreement PZ00P2 223297. O.M. was supported by the fundamental junior IUF grant G752IUFMAR.

\section{Topological Recursion for the strip geometries}\label{sec:TR}

This section will give some basic background on Topological Recursion and recent developments within TR which make it possible to derive the free energy explicitly via TR. We want to highlight that the actual derivation will not be carried out by the recursive definition of TR, but rather via a universal duality, the so-called $x$--$y$ duality, which enables us to compute the free energy efficiently.

\subsection{Background on Topological Recursion}
TR is a recursive procedure which generates from the initial data of a spectral curve an infinite family of multidifferentials $\omega_{g,n}$ on $n$ copies of the spectral curve. The definition \cite{Eynard:2007kz} of $\omega_{g,n}$ is recursive in $2g+n-2$ with $g,n\in \mathbb{Z}_{\geq0}$ and has a very beautiful pictorial interpretation. We refer to \cite{Eynard:2014zxa} for a classical overview and \cite{Bouchard:2024fih} for a more recent one that focuses on algebraic structures.

The spectral curve, the initial data, is a tuple $(\Sigma,x,y,B)$ with $\Sigma$ a Riemann surface with two functions $x,y:\Sigma\to \mathbb{C}$. $B$ is a symmetric bidifferential on $\Sigma\times \Sigma$ with second order pole on the diagonal, no residue and normalized along the $A$-cycles for a choice of a basis. Another description of the spectral curve typically used in the literature is given in terms of an affine curve, the vanishing locus of a polynomial equation in $x,y$, that is 
\begin{align}\label{spectralcurveP}
    P(x,y)=0.
\end{align}
Actually, it is very common to relax this to general (non-algebraic) equations allowing for instance exponentials which would for instance correspond to logarithmic poles for $x$ and/or $y$ on $\Sigma$. 

Assume $x,y$ have disjoint sets of ramification points, and $x$ has simple ramification points. Topological recursion which defines all $\omega_{g,n}$ is then defined by the following recursive formula: \\
From the initial data, define $\omega_{0,1}=y\,dx$ and $\omega_{0,2}=B$. Then, for all $2g+n-1>0$, $\omega_{g,n+1}$ is given by 
\begin{align}
  \label{eq:TR-intro}
&  \omega_{g,n+1}(I,z)
  :=\sum_{p_i\in Ram(x)}
  \Res\displaylimits_{q\to p_i}
  K_i(z,q)\bigg(
   \omega_{g-1,n+2}(I, q,\sigma_i(q))\\
  &\qquad \qquad\qquad\qquad\qquad\qquad
  +
   \sum_{\substack{g_1+g_2=g\\ I_1\sqcup I_2=I\\
            (g_i,I_i)\neq (0,\emptyset)}}
   \omega_{g_1,|I_1|+1}(I_1,q)
  \omega_{g_2,|I_2|+1}(I_2,\sigma_i(q))\!\bigg)\nonumber%\\\nonumber
  %&\qquad \qquad+\delta_{n,0}\sum_{i=1}^M\Res\displaylimits_{q\to a_i}\int_{a_i}^q \omega_{0,2}(z, \bullet)[\hbar^{2g}]\left(\frac{1}{\alpha_i S(\alpha_i \hbar \partial_{x(q)})}\log (q-a_i)\right)dx(q).
\end{align}
Here, the recursion kernel is 
\[
K_i(z,q)=\frac{\frac{1}{2}\int^{q}_{\sigma_i(q)}\omega_{0,2}(z,\bullet)}{\omega_{0,1}(q)-\omega_{0,1}(\sigma_i(q))}.
\]
With $I=\{z_1,\ldots,z_n\}$, we denote the set of coordinates on the different copies of $\Sigma^n$. The set $Ram(x)$ is the set of ramification points of $x$, that is $p_i\in Ram(x)$ if $dx(p_i)=0$. At a (simple) ramification point $p_i \in Ram(x)$, there is a unique deck transformation defined by $\sigma_i$ with $x(q)=x(\sigma_i(q))$ and with $p_i$ as the fixed point, i.e. $\sigma_i(p_i)=p_i$.

The free energy will be denoted by $F_g=\omega_{g,0}$ which is not directly generated by \eqref{eq:TR-intro}, but defined by
\begin{align}\label{freeenergydefinition}
    F_g=\frac{1}{2-2g}\sum_{p_i\in Ram(x)}
  \Res\displaylimits_{q\to p_i} \omega_{g,1}(q)\Phi(q)
\end{align}
for $g\geq 2$ with $d\Phi(q)=\omega_{0,1}(q)$. For $g=0,1$, the definition is more involved and will not play any role here, see \cite{Eynard:2007kz} for details. The definition of $F_g=\omega_{g,0}$ arises from a more general principle valid in TR going from $\omega_{g,n+1}$ to $\omega_{g,n}$  by
\begin{align}
    \omega_{g,n}(I)=\frac{1}{2-2g-n}\sum_{p_i\in Ram(x)}
  \Res\displaylimits_{q\to p_i} \omega_{g,1+n}(q,I)\Phi(q).
\end{align}
For instance to derive $F_2$, the knowledge of $\omega_{2,1}$ is needed which further is derived from $\omega_{1,2}$ and $\omega_{1,1}$, where $\omega_{0,3}$ is needed to get $\omega_{1,2}$. Thus, TR is a rather tedious algorithm to derive for instance $F_g$ for large $g$.

From the definition \eqref{eq:TR-intro}, several properties follow more or less directly \cite{Eynard:2007kz}. For $2g+n-2>0$, all $\omega_{g,n}$:\begin{itemize}
    \item are symmetric 
    \item have poles only at the ramification points of $x$, that is at $p_i\in Ram(x)$
    \item are residue-free, that is $\Res\displaylimits_{q\to p_i}\omega_{g,n}(q,I)=0$ for all $p_i\in Ram(x)$
    \item are homogeneous, that is, under the rescaling $\omega_{0,1}\to \lambda \omega_{0,1}$ we get
    $\omega_{g,n}\to \lambda^{2-2g-n}\omega_{g,n}$
    \item are invariant under the transformations 
    \begin{align*}
        (x,y)&\to (x,y+R(x))\qquad \text{for some rational function $R$}\\
        (x,y)&\to \left(\frac{ax+b}{cx+d},y\, \frac{(cx+d)^2}{ad-bc}\right).
    \end{align*}
\end{itemize}
The last invariance property gives a hint that the multidifferentials are actually invariants for symplectic structures. The formal symplectic form $dx\wedge dy$ is invariant under the two transformations above. Thus, $\omega_{g,n}$ stays invariant under at least a subgroup of symplectomorphisms $dx\wedge dy\mapsto d\tilde{x}\wedge d\tilde{y}$. 

However, it is definitely not invariant under all, since swapping $x$ and $y$ leaves the symplectic form $dx\wedge dy$ (up to an overall factor of -1) invariant, but generates a completely different family of multidifferentials, which we will denote by $\omega_{g,n}^\vee$. This family has for instance poles at the ramification points of $y$ (rather than $x$) which were assumed to be different. Originally, the free energy $F_g$ was conjectured to be invariant under the $x$--$y$ swap \cite{Eynard:2007kz,Eynard_2008} but shortly later counter-examples were found \cite{Bouchard:2011ya}. In \cite{Eynard:2013csa}, it was further investigated that the difference between $F_g$ and the dual $F_g^\vee$ depends on the singular points of the curve, which were understood partially for meromorphic $x,y$. 

Recently, explicit formulas between the two families $\omega_{g,n}$ of the spectral curve $(\Sigma,x,y,B)$ and $\omega_{g,n}^\vee$ of the curve $(\Sigma,y,x,B)$ were first conjectured \cite{Borot:2021thu}, then proven for $g=0$ \cite{Hock:2022wer} and in general in \cite{Alexandrov:2022ydc}, see also \cite{Hock:2022pbw} for equivalent formulations. 

The explicit formula itself between the two families $\omega_{g,n}$ and $\omega_{g,n}^\vee$ will not play a role; rather, its implications will.

The general form is
\begin{align}\label{dualxyexpression}
    \omega_{g,n} = \text{Expr}_{g,n}\left(\big\{\omega_{h,m}^\vee\big\}_{2h+m-2 \leq 2g+n-2}, \{dy_i, dx_i\}_{i=1,\dots,n}, \bigg\{\frac{dy_i\,dy_j}{(y_i - y_j)^2}\bigg\}_{i,j=1,\dots,n}\right),
\end{align}
where $\text{Expr}_{g,n}$ is an algebraic combinatorial expression depending on $\omega_{h,m}^\vee$, the differentials $dx_i$, $dy_i$, and the regularizing term $\frac{dy_i\,dy_j}{(y_i - y_j)^2}$, which serves to regularize $\omega_{0,2}^\vee$ at the diagonal.

We will not write out the formula, since it consists of an intricate sum over graphs weighted by $\omega_{g',n'}^\vee$ and differential operators as indicated. For $n = 1$, the first two leading terms in \eqref{dualxyexpression} are of the form
\begin{align}\label{dualxyexpressionfirstterms}
    \omega_{g,1} &= -\omega_{g,1}^\vee + d\Bigg(\frac{\omega_{g-1,2}^\vee + \frac{1}{2}\sum_{\substack{g_1+g_2=g\\ g_i>0}} \omega_{g_1,1}^\vee\,\omega_{g_2,1}^\vee}{dx\,dy}\Bigg) +\sum_{m=2}^{3g-1}\left(d\frac{1}{dx}\right)^m \Omega_{g,m},
\end{align}
which will be of particular use later. Here, $\Omega_{g,m}$ is a one-form obtained from graphs decorated by $\omega_{g',n'}^\vee$ and derivatives thereof. We refer to \cite{Hock:2022pbw,Alexandrov:2022ydc} for the explicit expression. Note that $d\frac{1}{dx}$ first divides a differential form by $dx$ and then takes the exterior derivative, thus turns a one-form into a one-form. 

The formula \eqref{dualxyexpression} can actually be understood as a universal duality, independent of TR, but proven to hold for the TR-differentials $\omega_{g,n}$ for any spectral curve $(\Sigma, x, y, B)$ with \emph{meromorphic} $x, y$.

Several further developments building upon the theory of $x$--$y$ duality were achieved \cite{Alexandrov:2023jcj,Alexandrov:2023tgl,Alexandrov:2024ajj,Alexandrov:2024hgu,Alexandrov:2024tjo,Alexandrov:2024zku,Alexandrov:2025sap,Bouchard:2025rid,Hock:2023qii,Hock:2023dno,Hock:2025wlm,Hock:2025sxq}.

Most importantly for this article (since we are dealing with meromorphic $dx$ and $dy$ rather than meromorphic $x, y$) is the fact that the explicit $x$--$y$ duality formula \eqref{dualxyexpression} suggests an extension of TR that properly considers logarithmic poles of $x, y$ (i.e., simple poles of $dx, dy$). This new definition is called Log-TR, suggested in \cite{Hock:2023dno} and further refined in \cite{Alexandrov:2023tgl}.

Under certain circumstances (for instance, if a generic framing is taken for the mirror curve of a Calabi-Yau threefold) Log-TR reduces to TR, which motivated the celebrated remodeling conjecture \cite{Bouchard:2007ys}, later proved in \cite{Eynard:2012nj,Fang:2016svw}.

For fixed framing $f = 0$ or $f=-1$, the remodeling conjecture does not hold; see, for instance, \cite{Bouchard:2011ya} for a discussion. However, this situation is rescued by the setting of Log-TR, which again arises by enforcing the $x$--$y$ duality. Note further that Log-TR is compatible with some limits beyond the families studied in \cite{Borot:2023wik}.

We are mentioning Log-TR here because we want to apply implications of the $x$--$y$ duality which are valid in the context of mirror curves, where either the family $\omega_{g,n}$ or $\omega_{g,n}^\vee$ (or both) differs from the standard TR case.

Now, let us define Log-TR. First, denote the logarithmic poles of $y$ (which are not simultaneously poles of $dx$) by $a_1,\ldots,a_M$, and refer to them as log-vital singular points. The logarithmic poles of $y$ that are also poles of $dx$ can be neglected (this is exactly what happens for generic framing in the remodeling conjecture). At the points $a_1,\ldots,a_M$, the differential $dy$ has nonzero residues $\frac{1}{t_1},\ldots,\frac{1}{t_M}$, respectively.

Then, the multidifferentials of Log-TR are defined by $\omega_{0,1} = y\,dx$, $\omega_{0,2} = B$, and, for negative Euler characteristic $\chi = 2 - 2g - (n+1) < 0$, recursively by
\begin{align}
  \label{eq:TR-introLog}
  \omega_{g,n+1}(I,z)
  &:= \sum_{p_i \in {Ram}(x)}
  \Res\displaylimits_{q \to p_i}
  K_i(z,q)\bigg(
   \omega_{g-1,n+2}(I, q,\sigma_i(q))\\
  &\qquad\qquad\qquad\qquad
  +
   \sum_{\substack{g_1+g_2=g\\ I_1 \sqcup I_2 = I\\
            (g_i,I_i) \neq (0,\emptyset)}}
   \omega_{g_1,|I_1|+1}(I_1,q)
  \omega_{g_2,|I_2|+1}(I_2,\sigma_i(q))\bigg)\nonumber\\
  &+ \delta_{n,0}\sum_{i=1}^M\Res\displaylimits_{q \to a_i}\int_{a_i}^q \omega_{0,2}(z, \bullet)[\hbar^{2g}]\left(\frac{1}{t_i S(t_i \hbar \partial_{x(q)})} \log(q - a_i)\right) dx(q), \nonumber
\end{align}
with $S(t) = \frac{e^{t/2} - e^{-t/2}}{t}$.

We use the same symbols $\omega_{g,n}$ as in the definition of TR, since throughout the rest of this article we will refer to $\omega_{g,n}$ as defined via Log-TR. Only the $n=1$ sector is changed compared to TR, but this modification propagates recursively to all $\omega_{g,n}$.

For $2g + n - 2 > 0$, the differentials $\omega_{g,n}$ generated by Log-TR have poles only at the ramification points of $x$, and for $n = 1$, also at the log-vital singular points $a_i$. Furthermore, the $\omega_{g,n}$ are symmetric and residue-free. Note that even if $x$ is unramified, $\omega_{g,1}$ can still be non-trivial if log-vital singular points exist.

The $x$--$y$ duals, $\omega_{g,n}^\vee$, are defined by swapping $x$ and $y$. One then takes residues around $p_i^\vee \in {Ram}(y)$ and considers the dual log-vital singular points of $x$, denoted $a_1^\vee, \ldots, a_{M^\vee}^\vee$. The two families $\omega_{g,n}$ for the spectral curve $(\Sigma, x, y, B)$ and $\omega_{g,n}^\vee$ for $(\Sigma, y, x, B)$, both defined using Log-TR, are related via the expression \eqref{dualxyexpression}, see \cite{Alexandrov:2023tgl}.

\subsection{Deriving the Free Energy}
In this subsection, we will derive the free energy $F_g$ of the spectral curve \eqref{spectralcurveintro}
\begin{align*}
    A(e^x,e^y)=(1-e^y)\prod_{j=1}^s(1-\beta_j e^y)+(-1)^f e^xe^{y(1+f)} \prod_{j=1}^r(1-\alpha_j e^y)=0
\end{align*}
which can be parametrized by 
\begin{align}\nonumber
    x=&\log\left(\frac{(1-z) \underset{j=1}{\overset{s}{\prod}} (1-\beta_j z)}{ \underset{j=1}{\overset{r}{\prod}}(1-\alpha_j z)}\right)-(1+f)\log(-z) \\\label{parametrizedspectralcurve}
    y=&\log (z).
\end{align}
Note that we have defined $F_g$ in \eqref{freeenergydefinition} for the $\omega_{g,n}$ generated by TR rather than by Log-TR. Since logarithmic singularities appear in the spectral curve \eqref{parametrizedspectralcurve}, it is fair to ask for an adjustment of the definition of $F_g$ in the context of Log-TR. Due to the recent definition of Log-TR, this has not yet been proposed. However for the curve considered here \eqref{parametrizedspectralcurve}, this is actually not necessary since we are in the situation where Log-TR breaks down to TR. The logarithmic pole of $y$ at $z=0,\infty$ is also present for $x$ except if $f= -1$ or $s-r-f=0$. Thus we consider $f\notin\{-1,s-r\}$. The definition of $F_g$ from \eqref{freeenergydefinition} will therefore give the correct free energy.

Instead of using the actual algorithmic topological recursion which amounts to $2g-1$ recursive steps to compute $\omega_{g,1}$, we will apply the $x$--$y$ duality to derive all $F_g$ without applying any recursion. 
\begin{remark}
    In principle, the result for $F_g$ is known for the curve \eqref{spectralcurveintro}. One can deduce it via circumventing with the remodeling conjecture \cite{Bouchard:2007ys} to Gromov-Witten theory. It can be written in terms of Hodge class integrals on the moduli space of complex curves. Although the free energy $F_g$ was computed using the topological vertex method \cite{Iqbal:2004ne} a direct computation via Topological Recursion was never performed in the literature which  we provide in this article and as a bonus we also gain further insights. 
    
\end{remark}
Since the $x$--$y$ duality will play a fundamental role in the derivation, it is important to mention again that $\omega_{g,n}$ is still generated by TR whereas the dual correlators $\omega_{g,n}^\vee$ necessarily need Log-TR, since the $x$--$y$ duality only holds in the setting of Log-TR if logarithmic poles are present, even if one of the families breaks down to TR which is the case for $\omega_{g,n}$. For $\omega_{g,n}^\vee$, a simple situation appears since $y$ is unramified, the first two lines of (the $x$--$y$ dual of) \eqref{eq:TR-introLog} vanish and we are left with contributions from the logarithmic poles of $x$ which are not poles of $y$. Due to this formula, all $\omega_{g,n}^\vee$ with $2g+n-2>0$ can be written for the curve \eqref{parametrizedspectralcurve} as
\begin{align}\label{omgdual}
    \omega_{g,n}^\vee=\delta_{n,1}[\hbar^{2g}]dy \frac{1}{S(\hbar \partial_y)}x.
\end{align}
Recall $S(t)=\frac{e^{t/2}-e^{-t/2}}{t}$.

From \eqref{omgdual}, all poles of $\omega_{g,n}^\vee$ are just located at the log-vital points of $x$, which are at $z=1,\frac{1}{\alpha_j},\frac{1}{\beta_j}$. Following the definition of the dual free energy $F_g^\vee$ in the sense of \eqref{freeenergydefinition}, we actually find that $F_g^\vee=0$. 

In the setting, when two families $\omega_{g,n}$ and $\omega_{g,n}^\vee$ are related by \eqref{dualxyexpression} and  $F_g$ (and $F_g^\vee$) is defined by \eqref{freeenergydefinition} (and its dual version), an explicit formula for the difference $F_g-F_g^\vee$ is found in \cite[Thm. 1.1]{Hock:2025wlm}. This formula involves all $\Omega_{g,m}$ of \eqref{dualxyexpressionfirstterms}. Since $F_g^\vee=0$, we actually find a new formula for $F_g$ in terms of the dual correlators $\omega_{g',n'}^\vee$ instead of $\omega_{g,n}$. The new formula for the curve \eqref{spectralcurveintro} reads
\begin{align}\nonumber
    F_g=&F_g-F_g^\vee\\
    =&\frac{1}{2g-2}\sum_{p\in \{1,\frac{1}{\alpha_j},\frac{1}{\beta_j}\}}\Res\displaylimits_{q\to p} \sum_{m=2}^{3g-1}(-1)^{m-1}\frac{d^{m-1}y}{dx^{m-1}} \Omega_{g,m}.
\end{align}
In the vicinity of $z\in \{1,\frac{1}{\alpha_j},\frac{1}{\beta_j}\}$, we can integrate for each $m\geq 2$ exactly $m-1$ times by parts, since $y$ is regular at those points. (Note that we cannot integrate $m$ times, since $\int y\,dx $ has a logarithmic pole at $z\in \{1,\frac{1}{\alpha_j},\frac{1}{\beta_j}\}$). Thus we derive  
\begin{align}\label{fgstep1}
    F_g=\frac{1}{2-2g}\sum_{p\in \{1,\frac{1}{\alpha_j},\frac{1}{\beta_j}\}}\Res\displaylimits_{q\to p} y\sum_{m=2}^{3g-1}\left(d\frac{1}{dx}\right)^{m-1} \Omega_{g,m}.
\end{align}
In the next step, we observe that $\sum_{m=2}^{3g-1}\big(d\frac{1}{dx}\big)^{m-1} \Omega_{g,m}$ is an exact one-form (even for every $m$). The residue of an exact one-form vanishes at each of its poles, that gives 
\begin{align*}
    0=\Res\displaylimits_{q\to p} \sum_{m=2}^{3g-1}\left(d\frac{1}{dx}\right)^{m-1} \Omega_{g,m}\qquad \forall p\in \mathbb{C}.
\end{align*}
We can subtract this from \eqref{fgstep1} multiplied by $y(p)=\log (p)$ for any $p\in \{1,\frac{1}{\alpha_j},\frac{1}{\beta_j}\}$
\begin{align}\label{fgstep2}
    F_g=\frac{1}{2-2g}\sum_{p\in \{1,\frac{1}{\alpha_j},\frac{1}{\beta_j}\}}\Res\displaylimits_{q\to p} (y(q)-y(p))\sum_{m=2}^{3g-1}\left(d\frac{1}{dx}\right)^{m-1} \Omega_{g,m}.
\end{align}
In the last step, we replace after integration \eqref{dualxyexpressionfirstterms}
\begin{align}\label{fgstep3}
    \sum_{m=2}^{3g-1}\left(d\frac{1}{dx}\right)^{m-1} \Omega_{g,m}=\left(\int^q \omega_{g,1}+\int^q \omega_{g,1}^\vee-\frac{\omega_{g-1,2}^\vee + \frac{1}{2}  \underset{\substack{g_1+g_2=g\\ g_i>0}}{\sum} \omega_{g_1,1}^\vee\,\omega_{g_2,1}^\vee}{dx\,dy}\right) dx,
\end{align}
where the integration constant is irrelevant.
On the RHS, we have several terms, where each of them is not an exact one-form, but the whole RHS is still exact. However, due to the carefully chosen constant $y(p)$, we have  that $(y(q)-y(p))$ vanishes linearly in \eqref{fgstep2}. Thus, the first term on the RHS of \eqref{fgstep3} does not contribute, also $\omega_{g-1,2}^\vee=0$ for $g>0$. Finally, we end up with 
\begin{align}\label{fgcompute}
     F_g=\frac{1}{2-2g}\sum_{p\in \{1,\frac{1}{\alpha_j},\frac{1}{\beta_j}\}}\Res\displaylimits_{q\to p} (y(q)-y(p)) \left(\int^q \omega_{g,1}^\vee-\frac{ \frac{1}{2}\underset{\substack{g_1+g_2=g\\ g_i>0}}{\sum} \omega_{g_1,1}^\vee\,\omega_{g_2,1}^\vee}{dx\,dy}\right) dx.
\end{align}
Inserting the explicit expression of $\omega_{g,1}^\vee$ from \eqref{omgdual}, the final residue calculation reduces to some identities of polylogarithms. This is outsourced to Appendix \ref{appendixa}. To have a nicer representation of the results, we set $1=\beta_0$ for the first factor in \eqref{parametrizedspectralcurve}. 
By the computation in Appendix \ref{appendixa}, we derive the main result of the article:
\begin{main}
The spectral curve 
\begin{align*}\nonumber
    x=&\log\left(\frac{\prod_{j=0}^s (1-\beta_j z)}{\prod_{j=1}^r(1-\alpha_j z)}\right)-(1+f)\log(-z) \\
    y=&\log (z)
\end{align*}
computes by TR the free energy for $g\geq 2$
\begin{align}\nonumber
    F_g=&-\frac{1+r+ s}{2} \frac{B_{2g}B_{2g-2}}{ 2g (2-2g)\cdot (2g-2)!}-\sum_{1\leq i < j\leq r}\frac{B_{2g}\mathrm{Li}_{3-2g}(\frac{\alpha_i}{\alpha_j})}{2g (2g-2)!}\\\label{freeenergyresult}
    &-\sum_{0\leq i <  j\leq s}\frac{B_{2g}\mathrm{Li}_{3-2g}(\frac{\beta_i}{\beta_j})}{2g (2g-2)!}+\sum_{\substack{ 1\leq i\leq r \\ 0\leq j\leq s}}\frac{B_{2g}\mathrm{Li}_{3-2g}(\frac{\beta_i}{\alpha_j})}{2g (2g-2)!}.
\end{align}
\end{main}

Sending the Kähler parameters to zero just the first term survives with the prefactor $1+r+s$  corresponding to the Euler characteristic of the toric Calabi-Yau. Thus, this computation gives an independent proof for \cite[Conjecture 3]{Bouchard:2011ya} for strip geometries.

For $|\frac{\beta_i}{\beta_j}|<1$, any polylogarithm can be expanded as
\begin{align*}
    \mathrm{Li}_{3-2g}(\frac{\beta_i}{\beta_j})=\sum_{n=1}^\infty\frac{(\beta_i/\beta_j)^n}{n^{3-2g}}. 
\end{align*}
Interchanging the $\hbar$-sum and the sum over $n$, one can resum for any $n$ the sum over $g$ (formally),
\begin{align*}
   \frac{(\frac{\beta_i}{\beta_j})^n}{n} \sum_{g=1}^\infty\hbar^{2g-2}\frac{n^{2g-2}B_{2g}}{2g (2g-2)!}=\frac{(\frac{\beta_i}{\beta_j})^n}{n(e^{\frac{\hbar n}{2}}-e^{-\frac{\hbar n}{2}})^2}.
\end{align*}
In case of $\beta_i=\beta_j$, this produces exactly the double Bernoulli number term. The remaining sum over $n$ combines to the MacMahon function. Thus, we deduce (formally)
\begin{align}
\label{DTpartitionfn}
    Z=\exp\sum_g \hbar^{2g-2}F_g=\prod_{n=1}^\infty\frac{\prod_{i,j}(1-\frac{\alpha_i}{\beta_j}q^n )^{n}}{\prod_{i,j}(1-\frac{\beta_i}{\beta_j}q^n )^{n/2}\prod_{i,j}(1-\frac{\alpha_i}{\alpha_j}q^n )^{n/2}},
\end{align}
where the indices $(i,j)$ range from 0 to $s$ for $\beta$'s and $(i,j)$ range from 1 to $r$ for $\alpha$'s. This is the result in one sector. A proper mathematical derivation would be via resurgence applying Borel summation techniques from \cite{MR4394512} and the final result would depend on the integration contour \cite{Alim:2021mhp}.

\section{5D BPS index from strips}\label{sec:DT}
In this section, we compare the explicit results obtained from TR in \eqref{freeenergyresult} with the free energy of topological string theory \eqref{freeenergyintro} via the identification \eqref{omegaintro} of 5D BPS index. Our proposal amounts to computing the 5D BPS indices for strip geometries using the machinery of TR and the newly established $x$--$y$ duality \cite{Alexandrov:2022ydc,Hock:2022wer,Hock:2025wlm} together with Log-TR \cite{Hock:2023dno,Alexandrov:2023tgl}.

Starting from $Z_{\mathrm{top}}$, following \cite{Ooguri:2010yk}, one can compute the BPS partition function in the noncommutative chamber as 
\begin{equation}
Z_{\mathrm{BPS}} = Z_{\mathrm{top}} (q,Q) Z_{\mathrm{top}} (q,Q^{-1})
\end{equation}
for these geometries, where the BPS partition function was shown to agree with matrix model computations. More precisely, one should perform Borel resummation on the asymptotic series \eqref{freeenergyresult} in different sectors of the Borel plane following \cite{Grassi:2022zuk,Alim:2021mhp}. This way, one lands in one of the DT chambers \cite{Szendr_i_2008,Jafferis:2008uf,Ooguri:2010yk}. Writing $Z_{\mathrm{top}}$ as in \eqref{DTdef}, one can extract the DT invariants. In fact, these expressions are asymptotic in $\hbar$, but are convergent in $Q_i$. The convergence conditions on $Q_i$ determine the DT chamber, as was also observed in the computations of Nekrasov partition functions \cite{Nekrasov:2003rj,Nekrasov:2014nea}. In our expressions \eqref{freeenergyresult} and \eqref{DTpartitionfn}, we keep the parameters $\alpha_i$ and $\beta_i$ quite general, without specifying the exact combinations $Q_i$'s, the complex structure parameters that correspond to them. The right identification of the parameters is made once we consider the convergence of such series in $Q_i$'s. 

There are a few already established techniques for computing these 5D indices using other methods. One of them is the ``exponential networks'' \cite{Eager:2016yxd,Banerjee:2018syt,Banerjee:2019apt,Banerjee:2022oed,Banerjee:2024smk}. Briefly, the idea is to study the 3D-5D coupled system that we introduced in \textsection \ref{sec:intro}. Then one analyzes the wall-crossing of the 3D-5D BPS states. The information of the 5D BPS spectrum can be extracted by constructing a nonabelianization map for the exponential networks. In other words, the wall-crossing formula for 3D-5D BPS states, which is of the universal Kontsevich-Soibelman form \cite{kontsevich2008stabilitystructuresmotivicdonaldsonthomas}, one can extract 5D BPS indices, by looking at wall-crossing of some particular 3D-5D BPS states and this is achieved by analyzing the topologies of the exponential networks associated to the corresponding toric CY threefolds. There are several other techniques to compute these invariants, for example using the BPS quiver and studying the moduli space of stable representations therein \cite{Beaujard:2020sgs,Mozgovoy:2020has}. Moreover, there exist techniques for computing them using supersymmetric localization \cite{Nekrasov:2003rj,Awata:2008ed} or using  methods from matrix model \cite{Ooguri:2010yk,Sulkowski:2009rw}, gluing of topological vertex \cite{Iqbal:2004ne}, to name a few. The strip geometries are rather well-studied and many of the techniques mentioned above can extract the 5D BPS indices quite efficiently. Each of these methods brings forth certain aspects of mirror symmetry, such as the counting in \cite{Banerjee:2019apt,Banerjee:2020moh} are related to A-branes which are objects in Fukaya category, whereas the BPS quivers in \cite{Mozgovoy:2020has} are related to counting of B-branes which are objects in the derived category of coherent sheaves. This exemplifies one aspect of homological mirror symmetry. 

Advent of supersymmetric localization \cite{Nekrasov:2014nea,Nekrasov:2016qym,Jeong:2025yys} gave rise to yet another set of powerful techniques to study these 5D indices. In this situation, one needs to study certain moduli space which are multiplicative Higgs bundles. This is a more direct path towards quantization and as we shall see in an upcoming work \cite{Banerjee:2025shz} topological recursion is useful in that context too. Finally, there are several studies to understand the integrable structures of these 5D theories \cite{Nekrasov:2003af,Nekrasov:2004ws,Ooguri:2010yk}. 

The direct computation using TR has also its own implications. GW invariants can be reformulated using ``remodeling conjecture'' \cite{Bouchard:2007ys} on CY threefold $\mathcal{X}$.  The result $F=\sum_{g\geq 0} \hbar^{2g-2} F_g$ is an asymptotic series in $\hbar$,  where coefficients of each of the terms in $F_g$ in \eqref{freeenergyresult} encode the Gopakumar-Vafa invariants, for strip geometries. These GV invariants have clear interpretation in M-theory, as counting of M2-branes. The counts are also equal to the 5D BPS indices. In this work, we perform a verification of these statements through topological recursion. We provide some checks of this fact by comparing the free energy \eqref{freeenergyresult} and 5D invariants computed using other techniques, showing that they are in accordance with each other.\footnote{The computations of the 5D indices referred to in the literature are 
carried out on both the A--model and B--model sides of these geometries: on the A--model (Fukaya category) side one relies on the techniques of \emph{exponential networks}, while on the B--model (derived category of coherent sheaves) side one uses quiver descriptions and methods from representation theory. A thorough physical justification connecting both techniques is now available in the literature. In this sense, recovering these numbers directly from Topological Recursion can be interpreted as a ``proof via TR''.} One crucial simplification for the cases here is that, because there is no compact four cycle, the wall-crossing of the 5D BPS states is tractable and we can extract them by essentially analyzing one of the wall-crossing chambers.

\subsection{Four punctured spheres}
We start from the resolved conifold. In this case the mirror curve \eqref{spectralcurveintro} has only one parameter $\alpha_1 = Q$. The genus $g$ free energy reads as 
\begin{equation}
F_g = - \frac{B_{2g}B_{2g-2}}{2g(2-2g)(2g-2)!} + \frac{B_{2g}\mathrm{Li}_{3-2g}(Q)}{2g(2g-2)!}.
\end{equation}
Depending on whether $|Q|<1$ or $|Q|>1$, one can write the sum in terms of $\mathrm{Li}_{3-2g}(Q)$ or $\mathrm{Li}_{3-2g}(Q^{-1})$ which are equal. However, as expansions in $Q$, only one of them is convergent, given the above choice. This was also observed in \cite{Szendr_i_2008}, giving two distinct chambers in complex structure moduli space. In this case, we infer \footnote{The bar on charges denote the anti D2-brane charges. The D-brane charges are classified by K-theory by the even integral cohomology of $\mathcal{X}$, $\Gamma \subset H^{\textrm{even}}(\mathcal{X},\mathbb{Z})$. A general charge vector is written as $\gamma \in \Gamma$, where $\gamma = (p^0,p^a,q_a,q_0)$, where focusing on a D2-branes that wraps a curve $C$ of class $\beta$ particularly $\gamma_{\textrm{D2}} = (0,0, \beta, \frac12 \beta\cdot c_1(\mathcal{X}))$. Since for CY3, $c_1(\mathcal{X})=0$, one has $\gamma_{D2}= -\gamma_{\overline{D2}} = (0,0,\beta,0)$. For all the examples we use this notation with overline for the anti-D2 branes.}

\begin{equation}
    \begin{cases}
      \Omega(n D0) &= -2, n\geq 1;\\
      \Omega (D2_Q-k D0 ) &=  1, k\geq 0; \\
      \Omega (\overline{D2}_Q-k D0 )& =  1, k\geq 0 
    \end{cases}       
\end{equation}
This matches with the results of \cite{Ooguri:2010yk,Szendr_i_2008,Banerjee:2019apt}. \\

For the other resolution of $\mathbb{P}^1$, in \eqref{spectralcurveintro}, one has $\beta_1 = Q$ and the rest are all zero.

The free energy does not depend on the choice of basepoint. The result in this case is ($g\geq 2$)
\begin{equation}
\label{freeC3Z2}
F_g = -\frac{B_{2g}B_{2g-2}}{2g (2g-2) (2g-2)!}
+ (-1) \frac{B_{2g}}{2g(2g-2)!} \mathrm{Li}_{3-2g}(Q)
\end{equation}

\begin{equation}
    \begin{cases}
      \Omega(n D0) &= -2, n\geq 1;\\
      \Omega (D2_Q-k D0 ) &=  -1, k\geq 0; \\
      \Omega (\overline{D2}_Q-k D0 ) &=  -1, k\geq 0 
    \end{cases}       
\end{equation}

From this one can actually reconstruct the Donaldson-Thomas partition function 
\begin{equation}
Z_{\mathrm{DT}} = M(q)^{-2} \prod_{n=1}^\infty (1-Qq^n)^{-n}
\end{equation}
equation (4.36) in \cite{Banerjee:2019apt}. 

Next we go into the computation for more complicated toric trees obtained by gluing. 

\subsection{Examples of five punctured spheres}

For the example of the SPP geometry, one can compute
\begin{equation}
    \begin{cases}
      \Omega(n D0) &= -3, n\geq 1;\\
      \Omega (D2_{Q\mu}-k D0 ) &=  -1, k\geq 0; \\
      \Omega ({D2}_Q-k D0 ) &=  1, k\geq 0;\\
      \Omega ({D2}_\mu-k D0 ) &= 1, k\geq 0;\\
      \Omega (\overline{D2}_{Q\mu}-k D0 ) &=  -1, k\geq 0; \\
      \Omega (\overline{D2}_Q-k D0 ) &=  1, k\geq 0;\\
      \Omega (\overline{D2}_\mu-k D0 )& = 1, k\geq 0;
    \end{cases}       
\end{equation}

 Next, one could consider the case of gluing two copies of $\mathcal{O}(0,-2)$, corresponding to $s=2$ which corresponds to resolution of $\mathbb{C}\times\mathbb{C}^2/\mathbb{Z}_3$. In this case, as in \cite{Panfil:2018faz}, we consider $\beta_1 = Q_1,\beta_2 = Q_1Q_2$. Looking at the genus $g$ free energy \eqref{freeenergyresult}, one can check 
 \begin{equation}
    \begin{cases}
      \Omega(n D0) &= -3, n\geq 1;\\
      \Omega (D2_{Q_1Q_2}-k D0 ) &=  -1, k\geq 0; \\
      \Omega ({D2}_{Q_1}-k D0 ) &=  -1, k\geq 0;\\
      \Omega ({D2}_{Q_2}-k D0 ) &= -1, k\geq 0;\\
      \Omega (\overline{D2}_{Q_1Q_2}-k D0 ) &=  -1, k\geq 0; \\
      \Omega (\overline{D2}_{Q_1}-k D0 ) &=  -1, k\geq 0;\\
      \Omega (\overline{D2}_{Q_2}-k D0 ) &= -1, k\geq 0;
    \end{cases}       
\end{equation}
which matches the computations of \cite{MR3713014,Mozgovoy:2020has,MR3000467,MR3352820}.

\subsection{Examples of six punctured spheres}

We take the example of \cite{Panfil:2018faz} for the six-punctured sphere with $r=2, s=1$, with $\alpha_1 = Q_1, \alpha_2 = Q_1Q_2Q_3$ and $\beta_1= Q_1Q_2$. The 5D indices can then be read off from \eqref{freeenergyresult} as 
\begin{equation}
    \begin{cases}
      \Omega(n D0)& = -4, n\geq 1;\\
      \Omega (D2_{Q_1}-k D0 ) &=  1, k\geq 0; \\
      \Omega ({D2}_{Q_2}-k D0 ) &=  1, k\geq 0;\\
      \Omega ({D2}_{Q_3}-k D0 ) &= 1, k\geq 0;\\
      \Omega ({D2}_{Q_1Q_2Q_3}-k D0 ) &= 1, k\geq 0 ;\\
      \Omega ({D2}_{Q_2Q_3}-k D0 ) &= -1, k\geq 0 ;\\
      \Omega ({D2}_{Q_1Q_2}-k D0 ) &= -1, k\geq 0 ;\\
      \Omega (\overline{D2}_{Q_1}-k D0 ) &=  1, k\geq 0; \\
      \Omega (\overline{D2}_{Q_2}-k D0 ) &=  1, k\geq 0;\\
      \Omega (\overline{D2}_{Q_3}-k D0 ) &= 1, k\geq 0;\\
      \Omega (\overline{D2}_{Q_1Q_2Q_3}-k D0 ) &= 1, k\geq 0 ;\\
      \Omega (\overline{D2}_{Q_2Q_3}-k D0 ) &= -1, k\geq 0 ;\\
      \Omega (\overline{D2}_{Q_1Q_2}-k D0 ) &= -1, k\geq 0,
    \end{cases}       
\end{equation}
in agreement with \cite{Mozgovoy:2020has,MR3352820}.

\appendix 
\section{Some Identities of Polylogarithms}\label{appendixa}
Here, we complete the free energy computation by using well-known identities of polylogarithms. Recall \eqref{fgcompute}
\begin{align*}
     F_g=\frac{1}{2-2g}\Res\displaylimits_{q\to p\in \{\frac{1}{\alpha_i},\frac{1}{\beta_j}\}} (y(q)-y(p)) \left(\int^q \omega_{g,1}^\vee-\frac{ \frac{1}{2}\underset{\substack{g_1+g_2=g\\ g_i>0}}{\sum} \omega_{g_1,1}^\vee(q)\,\omega_{g_2,1}^\vee(q)}{dx(q)\,dy(q)}\right) dx(q)
\end{align*}
with 
\begin{align*}\nonumber
    x(z)=\log\bigg(\frac{\prod_{j=0}^s (1-\beta_j z)}{\prod_{j=1}^r(1-\alpha_j z)}\bigg)-(1+f)\log(-z),\qquad 
    y(z)=\log (z)
\end{align*}
and 
\begin{align*}
    \omega_{g,n}^\vee=&\delta_{n,1}[\hbar^{2g}]dy \frac{1}{S(\hbar \partial_y)}x\\
    =&\delta_{n,1}[\hbar^{2g}]\frac{dy}{S(\hbar)}\bigg(\sum_{j=1}^r \mathrm{Li}_{1-2g}(\alpha_j z)-\sum_{j=0}^s\mathrm{Li}_{1-2g}(\beta_j z)\bigg)
\end{align*}
where $S(t)=\frac{e^{t/2}-e^{-t/2}}{t}$. We have also used $\big(z\frac{d}{dz}\big)^{2g}\log(1-a z)=-\mathrm{Li}_{1-2g}(az)$, which follows directly from the definition of the polylogarithm.\\

We split the residue computation into two contributions: one coming from $\int^q \omega_{g,1}^\vee$ and the other from the quadratic term $\underset{\substack{g_1+g_2=g,\\ g_i>0}}{\sum} \omega_{g_1,1}^\vee(q)\,\omega_{g_2,1}^\vee(q)$.\\

For $p=\frac{1}{\beta_j}$, we find that the first contribution is
\begin{align}\nonumber
    &\Res\displaylimits_{q\to \frac{1}{\beta_j}}\log (q \beta_j)\,dx(q)\int^q \omega_{g,1}^\vee\\\nonumber
    =&\Res\displaylimits_{q\to \frac{1}{\beta_j}}\log (q \beta_j)\,dx(q)\,[\hbar^{2g}]\frac{1}{S(\hbar)}\bigg(\sum_{i} \mathrm{Li}_{2-2g}(\alpha_i q)-\sum_{i}\mathrm{Li}_{2-2g}(\beta_i q)\bigg)\\\nonumber
    =&-[\hbar^{2g}]\frac{1}{S(\hbar)}\Res\displaylimits_{q\to\frac{1}{\beta_j}}\log (q \beta_j)\,dq \bigg(\sum_i \frac{\beta_i}{\beta_i q-1}-\sum_i\frac{\alpha_i}{\alpha_i q-1}\bigg)\mathrm{Li}_{2-2g}(\beta_j q)\\\label{betaij}
    =&-[\hbar^{2g}]\frac{1}{S(\hbar)}\Res\displaylimits_{q\to 1}\log (q)\,dq \bigg(\sum_i \frac{\frac{\beta_i}{\beta_j}}{\frac{\beta_i}{\beta_j} q-1}-\sum_i\frac{\frac{\alpha_i}{\beta_j}}{\frac{\alpha_i}{\beta_j} q-1}\bigg)\mathrm{Li}_{2-2g}(q).
\end{align}

This residue can be derived from the series representation
\begin{align}\label{liexpansion}
    \mathrm{Li}_{-n}\left(e^{\mu}\right)=\frac{n!}{(-\mu)^{n+1}} -\sum_{k=0}^{\infty} \frac{B_{k+n+1}}{k! (k+n+1)}\mu^k \,\,,\, \, |\mu|\leq 2\pi.
\end{align}

Inserting this expansion with the change of variables $q=e^\mu$, we find
\begin{align*}
    &\Res\displaylimits_{q\to 1}\log (q)\,dq\,\frac{1}{a q-1}\mathrm{Li}_{2-2g}(q)
    =-\Res\displaylimits_{\mu\to 0}\mu e^\mu d\mu\, \frac{1}{a e^{\mu}-1}\frac{(2g-2)!}{\mu^{2g-1}}\\
    =&\frac{d^{2g-2}}{d\mu^{2g-2}}\frac{-\mu}{a-e^{-\mu}}=\frac{1}{a}(2g-2)\mathrm{Li}_{3-2g}(a).
\end{align*}

The last equality follows from the fact that $\frac{-\mu}{a-e^{-\mu}}$ is the exponential generating function of the polylogarithm at negative integers evaluated at $a$. For $a=1$, that is, when $\beta_i=\beta_j$, we must perform the computation separately since $\mathrm{Li}_{3-2g}(a)$ diverges at $a=1$. We find
\begin{align*}
    &\Res\displaylimits_{q\to 1}\log (q)\,dq\,\frac{1}{q-1}\mathrm{Li}_{2-2g}(q)=-B_{2g-2}.
\end{align*}

Thus, we can summarize:
\begin{align}\nonumber
    &\Res\displaylimits_{q\to \frac{1}{\beta_j}}\log (q \beta_j)\,dx(q)\int^q \omega_{g,1}^\vee\\\label{omg1f}
    =&[\hbar^{2g}]\frac{1}{S(\hbar)}B_{2g-2}-(2g-2)[\hbar^{2g}]\frac{1}{S(\hbar)}\bigg(\sum_{i\neq j} \mathrm{Li}_{3-2g}\left(\frac{\beta_i}{\beta_j}\right)-\sum_i\mathrm{Li}_{3-2g}\left(\frac{\alpha_i}{\beta_j}\right)\bigg).
\end{align}

At $p=\frac{1}{\alpha_j}$, the computation is the same, with the roles of $\alpha_j$ and $\beta_j$ exchanged.\\

Let us next compute the contribution where the integrand consists of the product $\omega_{g_1,1}^\vee \omega_{g_2,1}^\vee$ at $p=\frac{1}{\beta_j}$. We insert the expression for $\omega_{g',1}^\vee$. For the final residue, we need to compute (first for fixed $g_1+g_2=g$ but $g_i\neq 0$)
\begin{align*}
    &\Res\displaylimits_{q\to \frac{1}{\beta_j}}\log (q \beta_j)\,dy(q)\bigg(\sum_{i=1}^r \mathrm{Li}_{1-2g_1}(\alpha_i q)-\sum_{i=0}^s\mathrm{Li}_{1-2g_1}(\beta_i q)\bigg)\\
    & \times\bigg(\sum_{i=1}^r \mathrm{Li}_{1-2g_2}(\alpha_i q)-\sum_{i=0}^s\mathrm{Li}_{1-2g_2}(\beta_i q)\bigg)\\
    =&\Res\displaylimits_{q\to 1}\log (q)\,dy(q)\bigg(\sum_{i=1}^r \mathrm{Li}_{1-2g_1}\left(\frac{\alpha_i}{\beta_j} q\right)-\sum_{i=0}^s\mathrm{Li}_{1-2g_1}\left(\frac{\beta_i}{\beta_j} q\right)\bigg)\\
    &\times\bigg(\sum_{i=1}^r \mathrm{Li}_{1-2g_2}\left(\frac{\alpha_i}{\beta_j} q\right)-\sum_{i=0}^s\mathrm{Li}_{1-2g_2}\left(\frac{\beta_i}{\beta_j} q\right)\bigg)\\
    =&\Res\displaylimits_{\mu\to 0}\mu\, d\mu \bigg(\sum_{i=1}^r \mathrm{Li}_{1-2g_1}\left(\frac{\alpha_i}{\beta_j} e^\mu\right)-\sum_{i=0}^s\mathrm{Li}_{1-2g_1}\left(\frac{\beta_i}{\beta_j} e^\mu\right)\bigg)\\
    &\times\bigg(\sum_{i=1}^r \mathrm{Li}_{1-2g_2}\left(\frac{\alpha_i}{\beta_j} e^\mu\right)-\sum_{i=0}^s\mathrm{Li}_{1-2g_2}\left(\frac{\beta_i}{\beta_j} e^\mu\right)\bigg).
\end{align*}

We now have three contributions: either from $\beta_i = \beta_j$ in the first parenthesis, or from $\beta_i = \beta_j$ in the second parenthesis, or in both. The first two cases are symmetric under exchange of $g_1$ and $g_2$. The third is special and will be computed first.

Using the expansion \eqref{liexpansion} and considering only the terms with $\beta_i = \beta_j$ in both expressions, we obtain
\begin{align*}
    &\Res\displaylimits_{\mu\to 0}\mu\, d\mu\bigg(\frac{(2g_1-1)!}{\mu^{2g_1}}-\sum_{k=0}^\infty\frac{B_{k+2g_1}}{k! (2g_1+k)}\mu^k\bigg)\bigg(\frac{(2g_2-1)!}{\mu^{2g_2}}-\sum_{k=0}^\infty\frac{B_{k+2g_2}}{k! (2g_2+k)}\mu^k\bigg)\\
    =&-\frac{(2g_1-1)! B_{2g_1+2g_2-2}}{(2g_1-2)! (2g_1+2g_2-2)}-\frac{(2g_2-1)! B_{2g_1+2g_2-2}}{(2g_2-2)! (2g_1+2g_2-2)}=-B_{2g_1+2g_2-2}.
\end{align*}

Next, if for instance in the second parenthesis we consider $\beta_i \neq \beta_j$ (setting again $a = \frac{\beta_i}{\beta_j}$), the following computation arises:
\begin{align*}
    &\Res\displaylimits_{\mu\to 0}\mu\, d\mu\,\frac{(2g_1-1)!}{\mu^{2g_1}}\bigg(\frac{(2g_2-1)!}{(\mu+\log a)^{2g_2}}-\sum_{k=0}^\infty\frac{B_{k+2g_2}}{k! (2g_2+k)}(\mu+\log a )^k\bigg)\\
    =&\frac{(2g_1-1)!}{(2g_1-2)!}\frac{d^{2g_1-2}}{d\mu^{2g_1-2}}\bigg(\frac{(2g_2-1)!}{(\mu+\log a)^{2g_2}}-\sum_{k=0}^\infty\frac{B_{k+2g_2}}{k! (2g_2+k)}(\mu+\log a )^k\bigg)\bigg\vert_{\mu=0}\\
    =&(2g_1-1)\bigg(\frac{(2g_1+2g_2-3)! }{(\log a)^{2g_1+2g_2-2}}-\sum_{k=2g_1-2}^\infty\frac{B_{k+2g_2} }{(k-2g_1+2)! (2g_2+k)}(\log a )^{k-2g_1+2}\bigg)\\
    =&(2g_1-1)\bigg(\frac{(2g_1+2g_2-3)! }{(\log a)^{2g_1+2g_2-2}}-\sum_{k=0}^\infty\frac{B_{k+2g_2+2g_1-2} }{k! (2g_1+2g_2-2+k)}(\log a )^{k}\bigg)\\
    =&(2g_1-1)\mathrm{Li}_{3-2g_1-2g_2}(a),
\end{align*}
where the expansion \eqref{liexpansion} was applied in the last step.

Thus, for the residue at $p=\frac{1}{\beta_j}$, we summarize:
\begin{align}\nonumber
    &-\frac{1}{2}\Res\displaylimits_{q\to \frac{1}{\beta_j}} (y(q)-y(p))\,\frac{\underset{\substack{g_1+g_2=g,\\ g_i>0}}{\sum} \omega_{g_1,1}^\vee(q)\,\omega_{g_2,1}^\vee(q)}{dx(q)\,dy(q)}\, dx(q)\\\nonumber
    =&-\frac{1}{2}\sum_{\substack{g_1+g_2=g\\ g_i>0}}\bigg([\hbar^{2g_1}]\frac{1}{S(\hbar)}\bigg)\bigg([\hbar^{2g_2}]\frac{1}{S(\hbar)}\bigg)\bigg(-B_{2g-2}+(2g_1-1)\sum_{i\neq j}\mathrm{Li}_{3-2g}\left(\frac{\beta_i}{\beta_j}\right)\\\nonumber
    &-(2g_1-1)\sum_{i}\mathrm{Li}_{3-2g}\left(\frac{\alpha_i}{\beta_j}\right)
    +(2g_2-1)\sum_{i\neq j}\mathrm{Li}_{3-2g}\left(\frac{\beta_i}{\beta_j}\right)\\\nonumber
    &\qquad -(2g_2-1)\sum_{i}\mathrm{Li}_{3-2g}\left(\frac{\alpha_i}{\beta_j}\right)\bigg)\\\nonumber
    =&-\frac{1}{2}[\hbar^{2g}]\bigg(\frac{1}{S(\hbar)^2}-\frac{2}{S(\hbar)}\bigg)\bigg(-B_{2g-2}+(2g-2)\sum_{i\neq j}\mathrm{Li}_{3-2g}\left(\frac{\beta_i}{\beta_j}\right)\\\label{omg1f2}
    &\qquad -(2g-2)\sum_{i}\mathrm{Li}_{3-2g}\left(\frac{\alpha_i}{\beta_j}\right)\bigg).
\end{align}

For $p=\frac{1}{\alpha_j}$, we simply exchange all occurrences of $\beta$ with $\alpha$, and vice versa.\\

Adding up \eqref{omg1f} and \eqref{omg1f2}, we finally obtain the residue at $p=\frac{1}{\beta_j}$:
\begin{align*}
    &\Res\displaylimits_{q\to \frac{1}{\beta_j}} (y(q)-y(p)) \bigg(\int^q \omega_{g,1}^\vee - \frac{ \frac{1}{2}\underset{\substack{g_1+g_2=g,\\ g_i>0}}{\sum} \omega_{g_1,1}^\vee(q)\,\omega_{g_2,1}^\vee(q)}{dx(q)\,dy(q)}\bigg) dx(q)\\
    =&\frac{1}{2}[\hbar^{2g}]\frac{1}{S(\hbar)^2}\bigg(B_{2g-2} - (2g-2)\sum_{i\neq j} \mathrm{Li}_{3-2g}\left(\frac{\beta_i}{\beta_j}\right) + (2g-2)\sum_i \mathrm{Li}_{3-2g}\left(\frac{\alpha_i}{\beta_j}\right)\bigg).
\end{align*}

Last but not least, the expansion  
\[
\frac{1}{S(\hbar)^2} =1- \sum_{g>0}\hbar^{2g} \frac{B_{2g}}{(2g-2)! \, 2g}
\]
and the summation over all $j$, together with the residues at all $p = \frac{1}{\alpha_j}$ and the antisymmetry of the polylogarithm $\mathrm{Li}_{3-2g}(z)=\mathrm{Li}_{3-2g}(1/z)$, complete the derivation of all $F_g$ in \eqref{freeenergyresult}.

\section*{conflict of interest statement}
On behalf of all authors, the corresponding author states that there is no conflict of interest.

\bibliographystyle{halpha-abbrv}
\bibliography{omega.bib}

@article{Panfil:2018faz,
    author = "Panfil, Mi\l{}osz and Su\l{}kowski, Piotr",
    title = "{Topological strings, strips and quivers}",
    eprint = "1811.03556",
    archivePrefix = "arXiv",
    primaryClass = "hep-th",
    reportNumber = "CALT-2018-047",
    doi = "10.1007/JHEP01(2019)124",
    journal = "JHEP",
    volume = "01",
    pages = "124",
    year = "2019"
}

@article{Bouchard:2024fih,
    author = "Bouchard, Vincent",
    title = "{Les Houches lecture notes on topological recursion}",
    eprint = "2409.06657",
    archivePrefix = "arXiv",
    primaryClass = "math-ph",
    month = "9",
    year = "2024"
}

@article {MR3352820,
    AUTHOR = {Morrison, Andrew and Nagao, Kentaro},
     TITLE = {Motivic {D}onaldson-{T}homas invariants of small crepant
              resolutions},
   JOURNAL = {Algebra Number Theory},
  FJOURNAL = {Algebra \& Number Theory},
    VOLUME = {9},
      YEAR = {2015},
    NUMBER = {4},
     PAGES = {767--813},
      ISSN = {1937-0652,1944-7833},
   MRCLASS = {14N35 (14D20 14F05 14J32)},
  MRNUMBER = {3352820},
MRREVIEWER = {Si\ Li},
eprint = "1110.5976",
    archivePrefix = "arXiv",
    primaryClass = "math.AG",
    year = "2011",
       DOI = {10.2140/ant.2015.9.767},
       URL = {https://doi.org/10.2140/ant.2015.9.767},
}

@article{Eynard:2014zxa,
    author = "Eynard, B",
    title = "{A short overview of the ``Topological recursion''}",
    eprint = "1412.3286",
    archivePrefix = "arXiv",
    primaryClass = "math-ph",
    reportNumber = "IPHT-T14-033-CRM3335",
    month = "12",
    year = "2014"
}

@article{Eynard:2007kz,
    author = "Eynard, Bertrand and Orantin, Nicolas",
    title = "{Invariants of algebraic curves and topological expansion}",
    eprint = "math-ph/0702045",
    archivePrefix = "arXiv",
    reportNumber = "SPHT-07-021",
    doi = "10.4310/CNTP.2007.v1.n2.a4",
    journal = "Commun. Num. Theor. Phys.",
    volume = "1",
    pages = "347--452",
    year = "2007"
}

@article{Banerjee:2019apt,
    author = "Banerjee, Sibasish and Longhi, Pietro and Romo, Mauricio",
    title = "{Exponential BPS Graphs and D Brane Counting on Toric Calabi-Yau Threefolds: Part I}",
    eprint = "1910.05296",
    archivePrefix = "arXiv",
    primaryClass = "hep-th",
    doi = "10.1007/s00220-021-04242-4",
    journal = "Commun. Math. Phys.",
    volume = "388",
    number = "2",
    pages = "893--945",
    year = "2021"
}

@article{Borot:2021thu,
    author = {Borot, Ga{\"e}tan and Charbonnier, S{\'e}verin and Garcia-Failde, Elba and Leid, Felix and Shadrin, Sergey},
    title = "{Functional relations for higher-order free cumulants}",
    eprint = "2112.12184",
    archivePrefix = "arXiv",
    primaryClass = "math.OA",
    month = "12",
    year = "2021"
}

@article{Hock:2022wer,
    author = "Hock, Alexander",
    title = "{On the $x$--$y$ Symmetry of Correlators in Topological Recursion via Loop Insertion Operator}",
    eprint = "2201.05357",
    archivePrefix = "arXiv",
    primaryClass = "math-ph",
    doi = "10.1007/s00220-024-05043-1",
    journal = "Commun. Math. Phys.",
    volume = "405",
    number = "7",
    pages = "166",
    year = "2024"
}

@article{Alexandrov:2022ydc,
    author = "Alexandrov, Alexander and Bychkov, Boris and Dunin-Barkowski, Petr and Kazarian, Maxim and Shadrin, Sergey",
    title = "{A universal formula for the $x$--$y$ swap in topological recursion}",
    eprint = "2212.00320",
    archivePrefix = "arXiv",
    primaryClass = "math-ph",
    doi = "10.4171/JEMS/1615",
    month = "12",
    year = "2022"
}

@article{Hock:2022pbw,
    author = "Hock, Alexander",
    title = "{A simple formula for the $x$--$y$ symplectic transformation in topological recursion}",
    eprint = "2211.08917",
    archivePrefix = "arXiv",
    primaryClass = "math-ph",
    doi = "10.1016/j.geomphys.2023.105027",
    journal = "J. Geom. Phys.",
    volume = "194",
    pages = "105027",
    year = "2023"
}

@article{Alexandrov:2023jcj,
    author = "Alexandrov, Alexander and Bychkov, Boris and Dunin-Barkowski, Petr and Kazarian, Maxim and Shadrin, Sergey",
    title = "{KP integrability through the $x$--$y$ swap relation}",
    eprint = "2309.12176",
    archivePrefix = "arXiv",
    primaryClass = "math-ph",
    doi = "10.1007/s00029-025-01035-8",
    journal = "Selecta Math.",
    volume = "31",
    number = "2",
    pages = "42",
    year = "2025"
}

@article{Alexandrov:2023tgl,
    author = "Alexandrov, Alexander and Bychkov, Boris and Dunin-Barkowski, Petr and Kazarian, Maxim and Shadrin, Sergey",
    title = "{Log Topological Recursion Through the Prism of $x$--$y$ Swap}",
    eprint = "2312.16950",
    archivePrefix = "arXiv",
    primaryClass = "math-ph",
    doi = "10.1093/imrn/rnae213",
    journal = "Int. Math. Res. Not.",
    volume = "2024",
    number = "21",
    pages = "13461--13487",
    year = "2024"
}

@article{Alexandrov:2024ajj,
    author = "Alexandrov, Alexander and Bychkov, Boris and Dunin-Barkowski, Petr and Kazarian, Maxim and Shadrin, Sergey",
    title = "{Symplectic duality via log topological recursion}",
    eprint = "2405.10720",
    archivePrefix = "arXiv",
    primaryClass = "math-ph",
    doi = "10.4310/cntp.241203001416",
    journal = "Commun. Num. Theor. Phys.",
    volume = "18",
    number = "4",
    pages = "795--841",
    year = "2024"
}

@article{Alexandrov:2024hgu,
    author = "Alexandrov, Alexander and Bychkov, Boris and Dunin-Barkowski, Petr and Kazarian, Maxim and Shadrin, Sergey",
    title = "{Any topological recursion on a rational spectral curve is KP integrable}",
    eprint = "2406.07391",
    archivePrefix = "arXiv",
    primaryClass = "math-ph",
    month = "6",
    year = "2024"
}

@article{Alexandrov:2024tjo,
    author = "Alexandrov, Alexander and Bychkov, Boris and Dunin-Barkowski, Petr and Kazarian, Maxim and Shadrin, Sergey",
    title = "{Degenerate and Irregular Topological Recursion}",
    eprint = "2408.02608",
    archivePrefix = "arXiv",
    primaryClass = "math-ph",
    reportNumber = "MPIM-Bonn-2024",
    doi = "10.1007/s00220-025-05274-w",
    journal = "Commun. Math. Phys.",
    volume = "406",
    number = "5",
    pages = "94",
    year = "2025"
}

@article{Alexandrov:2024zku,
    author = "Alexandrov, Alexander and Bychkov, Boris and Dunin-Barkowski, Petr and Kazarian, Maxim and Shadrin, Sergey",
    title = "{KP integrability of non-perturbative differentials}",
    eprint = "2412.18592",
    archivePrefix = "arXiv",
    primaryClass = "math-ph",
    month = "12",
    year = "2024"
}

@article{Alexandrov:2025sap,
    author = "Alexandrov, Alexander and Bychkov, Boris and Dunin-Barkowski, Petr and Kazarian, Maxim and Shadrin, Sergey",
    title = "{Blobbed topological recursion and KP integrability}",
    eprint = "2505.03545",
    archivePrefix = "arXiv",
    primaryClass = "math-ph",
    month = "5",
    year = "2025"
}

@article{Bouchard:2025rid,
    author = "Bouchard, Vincent and Chidambaram, Nitin K. and Giacchetto, Alessandro and Shadrin, Sergey",
    title = "{Theta classes: generalized topological recursion, integrability and $\mathcal{W}$-constraints}",
    eprint = "2505.11291",
    archivePrefix = "arXiv",
    primaryClass = "math.AG",
    month = "5",
    year = "2025"
}

@article{Hock:2023qii,
    author = "Hock, Alexander",
    title = "{Laplace transform of the $x$--$y$ symplectic transformation formula in Topological Recursion}",
    eprint = "2304.03032",
    archivePrefix = "arXiv",
    primaryClass = "math-ph",
    doi = "10.4310/CNTP.2023.v17.n4.a1",
    journal = "Commun. Num. Theor. Phys.",
    volume = "17",
    number = "4",
    pages = "821--845",
    year = "2023"
}

@article{Hock:2023dno,
    author = "Hock, Alexander",
    title = "{$x$--$y$ duality in topological recursion for exponential variables via quantum dilogarithm}",
    eprint = "2311.11761",
    archivePrefix = "arXiv",
    primaryClass = "math-ph",
    doi = "10.21468/SciPostPhys.17.2.065",
    journal = "SciPost Phys.",
    volume = "17",
    number = "2",
    pages = "065",
    year = "2024"
}

@article{Hock:2025wlm,
    author = "Hock, Alexander",
    title = "{Symplectic (Non-)invariance of the Free Energy in Topological Recursion}",
    eprint = "2502.18115",
    archivePrefix = "arXiv",
    primaryClass = "math-ph",
    doi = "10.1007/s00220-025-05373-8",
    journal = "Commun. Math. Phys.",
    volume = "406",
    number = "8",
    pages = "192",
    year = "2025"
}

@article{Hock:2025sxq,
    author = "Hock, Alexander and Shadrin, Sergey",
    title = "{Quantum Curves in the Context of Symplectic Duality}",
    eprint = "2504.14924",
    archivePrefix = "arXiv",
    primaryClass = "math-ph",
    month = "4",
    year = "2025"
}

@article{Eynard_2008,
doi = {10.1088/1751-8113/41/1/015203},
year = {2007},
publisher = {},
volume = {41},
number = {1},
pages = {015203},
author = {Eynard, B and Orantin, N},
title = {{Topological expansion of mixed correlations in the Hermitian 2-matrix model and $x–y$ symmetry of the $F_g$ algebraic invariants}},
fjournal = {Journal of Physics A: Mathematical and Theoretical},
journal= {J. Phys. A: Math. Theor.}
}

@article{Bouchard:2007ys,
    author = "Bouchard, Vincent and Klemm, Albrecht and Mari\~{n}o, Marcos and Pasquetti, Sara",
    title = "{Remodeling the B-model}",
    eprint = "0709.1453",
    archivePrefix = "arXiv",
    primaryClass = "hep-th",
    reportNumber = "BONN-TH-2007-07, CERN-PH-TH-2007-153, NEIP-07-03",
    doi = "10.1007/s00220-008-0620-4",
    journal = "Commun. Math. Phys.",
    volume = "287",
    pages = "117--178",
    year = "2009"
}

@article{Eynard:2013csa,
    author = "Eynard, B. and Orantin, N.",
    title = "{About the $x$--$y$ symmetry of the $F_g$ algebraic invariants}",
    eprint = "1311.4993",
    archivePrefix = "arXiv",
    primaryClass = "math-ph",
    reportNumber = "IPHT-T13-252, CRM-3330",
    month = "11",
    year = "2013"
}

@article{Bouchard:2011ya,
    author = "Bouchard, Vincent and Su\l{}kowski, Piotr",
    title = "{Topological recursion and mirror curves}",
    eprint = "1105.2052",
    archivePrefix = "arXiv",
    primaryClass = "hep-th",
    reportNumber = "CALT-68-2836",
    doi = "10.4310/ATMP.2012.v16.n5.a3",
    journal = "Adv. Theor. Math. Phys.",
    volume = "16",
    number = "5",
    pages = "1443--1483",
    year = "2012"
}

@article{Borot:2023wik,
    author = {Borot, Ga{\"e}tan and Bouchard, Vincent and Chidambaram, Nitin Kumar and Kramer, Reinier and Shadrin, Sergey},
    title = "{Taking limits in topological recursion}",
    eprint = "2309.01654",
    archivePrefix = "arXiv",
    primaryClass = "math.AG",
    month = "9",
    year = "2023"
}

@article{Eynard:2012nj,
    author = "Eynard, Bertrand and Orantin, Nicolas",
    title = "{Computation of Open Gromov{\textendash}Witten Invariants for Toric Calabi{\textendash}Yau 3-Folds by Topological Recursion, a Proof of the BKMP Conjecture}",
    eprint = "1205.1103",
    archivePrefix = "arXiv",
    primaryClass = "math-ph",
    reportNumber = "IPHT-T12-030",
    doi = "10.1007/s00220-015-2361-5",
    journal = "Commun. Math. Phys.",
    volume = "337",
    number = "2",
    pages = "483--567",
    year = "2015"
}

@article{Forbes:2005xt,
    author = "Forbes, Brian and Jinzenji, Masao",
    title = "{Extending the Picard-Fuchs system of local mirror symmetry}",
    eprint = "hep-th/0503098",
    archivePrefix = "arXiv",
    doi = "10.1063/1.1996441",
    journal = "J. Math. Phys.",
    volume = "46",
    pages = "082302",
    year = "2005"
}

@article{Dijkgraaf:2006um,
    author = "Dijkgraaf, Robbert and Vafa, Cumrun and Verlinde, Erik",
    title = "{M-theory and a topological string duality}",
    eprint = "hep-th/0602087",
    archivePrefix = "arXiv",
    reportNumber = "HUTP-06-A001, ITFA-2006-05",
    month = "2",
    year = "2006"
}

@inproceedings{kontsevich1995homological,
  title={Homological algebra of mirror symmetry},
  author={Kontsevich, Maxim},
  booktitle={Proceedings of the international congress of mathematicians},
  pages={120--139},
  year={1995},
  organization={Springer}
}

@inproceedings{Kontsevich:2000yf,
    author = "Kontsevich, Maxim and Soibelman, Yan",
    title = "{Homological mirror symmetry and torus fibrations}",
    booktitle = "{KIAS Annual International Conference on Symplectic Geometry 
and Mirror Symmetry}",
    eprint = "math/0011041",
    archivePrefix = "arXiv",
    pages = "203--263",
    month = "11",
    year = "2000"
}

@article{Nekrasov:2014nea,
    author = "Nekrasov, Nikita and Okounkov, Andrei",
    title = "{Membranes and sheaves.}",
    eprint = "1404.2323",
    archivePrefix = "arXiv",
    primaryClass = "math.AG",
    doi = "10.14231/AG-2016-015",
    journal = "Algebr.  Geom.",
    volume = "3",
    number = "3",
    pages = "320--369",
    year = "2016"
}

@incollection {MR1363062,
    AUTHOR = {Kontsevich, Maxim},
     TITLE = {Enumeration of rational curves via torus actions},
 BOOKTITLE = {The moduli space of curves},
    SERIES = {Progr. Math.},
    VOLUME = {129},
     PAGES = {335--368},
 PUBLISHER = {Birkh\"auser Boston},
      YEAR = {1995},
      ISBN = {0-8176-3784-2},
   MRCLASS = {14N10 (14D22 14L30)},
  MRNUMBER = {1363062},
MRREVIEWER = {Anatoly\ Libgober},
eprint={hep-th/9405035},
      archivePrefix={arXiv},
      primaryClass={hep-th},
       DOI = {10.1007/978-1-4612-4264-2\_12},
       URL = {https://doi.org/10.1007/978-1-4612-4264-2_12},
}

@article{pandharipande2003questionsgromovwittentheory,
      title="{Three questions in Gromov-Witten theory}", 
      author={R. Pandharipande},
      year={2003},
      eprint={math/0302077},
      archivePrefix={arXiv},
      primaryClass={math.AG},
}

@article{Hori:2000kt,
    author = "Hori, Kentaro and Vafa, Cumrun",
    title = "{Mirror symmetry}",
    eprint = "hep-th/0002222",
    archivePrefix = "arXiv",
    reportNumber = "HUTP-00-A005",
    month = "2",
    year = "2000"
}

@article{Iqbal:2004ne,
    author = "Iqbal, Amer and Kashani-Poor, Amir-Kian",
    title = "{The Vertex on a strip}",
    eprint = "hep-th/0410174",
    archivePrefix = "arXiv",
    reportNumber = "SMS-0402, SLAC-PUB-10804, SU-ITP-04-39",
    doi = "10.4310/ATMP.2006.v10.n3.a2",
    journal = "Adv. Theor. Math. Phys.",
    volume = "10",
    number = "3",
    pages = "317--343",
    year = "2006"
}

@article{Aganagic:2000gs,
    author = "Aganagic, Mina and Vafa, Cumrun",
    title = "{Mirror symmetry, D-branes and counting holomorphic discs}",
    eprint = "hep-th/0012041",
    archivePrefix = "arXiv",
    reportNumber = "HUTP-00-A047",
    month = "12",
    year = "2000"
}

@article{Aganagic:2001nx,
    author = "Aganagic, Mina and Klemm, Albrecht and Vafa, Cumrun",
    title = "{Disk instantons, mirror symmetry and the duality web}",
    eprint = "hep-th/0105045",
    archivePrefix = "arXiv",
    reportNumber = "HUTP-01-A023, HU-EP-01-21",
    doi = "10.1515/zna-2002-1-201",
    journal = "Z. Naturforsch. A",
    volume = "57",
    pages = "1--28",
    year = "2002"
}

@article{Banerjee:2018syt,
    author = "Banerjee, Sibasish and Longhi, Pietro and Romo, Mauricio",
    title = "{Exploring 5d BPS Spectra with Exponential Networks}",
    eprint = "1811.02875",
    archivePrefix = "arXiv",
    primaryClass = "hep-th",
    doi = "10.1007/s00023-019-00851-x",
    journal = "Annales Henri Poincaré",
    volume = "20",
    number = "12",
    pages = "4055--4162",
    year = "2019"
}

@article{Banerjee:2020moh,
    author = "Banerjee, Sibasish and Longhi, Pietro and Romo, Mauricio",
    title = "{Exponential BPS graphs and D-brane counting on toric Calabi-Yau threefolds: Part II}",
    eprint = "2012.09769",
    archivePrefix = "arXiv",
    primaryClass = "hep-th",
    month = "12",
    year = "2020"
}

@article{Jeong:2025yys,
    author = "Jeong, Saebyeok and Lee, Norton",
    title = "{$Q$-operators, $q$-opers, and R-matrices in 5d $\mathcal{N}=1$ gauge theory}",
    eprint = "2507.15450",
    archivePrefix = "arXiv",
    primaryClass = "hep-th",
    reportNumber = "CERN-TH-2025-138",
    month = "7",
    year = "2025"
}

@incollection {MR4394512,
    AUTHOR = {Garoufalidis, Stavros and Kashaev, Rinat},
     TITLE = "{Resurgence of {F}addeev's quantum dilogarithm}",
 BOOKTITLE = {Topology and geometry---a collection of essays dedicated to
              {V}ladimir {G}. {T}uraev},
    VOLUME = {33},
     PAGES = {257--271},
 PUBLISHER = {Eur. Math. Soc.},
      YEAR = {2021},
      ISBN = {978-3-98547-001-3},
   MRCLASS = {57K16},
  MRNUMBER = {4394512},
       DOI = {10.4171/IRMA/33-1/14},
       URL = {https://doi.org/10.4171/IRMA/33-1/14},
}

@article{Iqbal:2007ii,
    author = "Iqbal, Amer and Kozcaz, Can and Vafa, Cumrun",
    title = "{The Refined topological vertex}",
    eprint = "hep-th/0701156",
    archivePrefix = "arXiv",
    doi = "10.1088/1126-6708/2009/10/069",
    journal = "JHEP",
    volume = "10",
    pages = "069",
    year = "2009"
}

@article{Banerjee:2026mwg,
    author = "Banerjee, Sibasish and Ishtiaque, Nafiz and Jeong, Saebyeok",
    title = "{Refined Invariants and Quantum Curves from Supersymmetric Localization}",
    eprint = "2601.07662",
    archivePrefix = "arXiv",
    primaryClass = "hep-th",
    month = "1",
    year = "2026"
}

@article{Banerjee:2025shz,
    author = "Banerjee, Sibasish and Hock, Alexander",
    title = "{Quantum Curve for strip geometries, Topological Recursion and open GW/DT invariants}",
    eprint = "2510.07146",
    archivePrefix = "arXiv",
    primaryClass = "math-ph",
    month = "10",
    year = "2025"
}

@article{Kidwai:2022kxx,
    author = "Kidwai, Omar and Osuga, Kento",
    title = "{Quantum curves from refined topological recursion: The genus 0 case}",
    eprint = "2204.12431",
    archivePrefix = "arXiv",
    primaryClass = "math.AG",
    doi = "10.1016/j.aim.2023.109253",
    journal = "Adv. Math.",
    volume = "432",
    pages = "109253",
    year = "2023"
}

@article{Gaiotto:2012rg,
    author = "Gaiotto, Davide and Moore, Gregory W. and Neitzke, Andrew",
    title = "{Spectral networks}",
    eprint = "1204.4824",
    archivePrefix = "arXiv",
    primaryClass = "hep-th",
    doi = "10.1007/s00023-013-0239-7",
    journal = "Annales Henri Poincaré",
    volume = "14",
    pages = "1643--1731",
    year = "2013"
}

@article{Banerjee:2022oed,
    author = "Banerjee, Sibasish and Longhi, Pietro and Romo, Mauricio",
    title = "{A-branes, Foliations and Localization}",
    eprint = "2201.12223",
    archivePrefix = "arXiv",
    primaryClass = "hep-th",
    doi = "10.1007/s00023-022-01231-8",
    journal = "Annales Henri Poincaré",
    volume = "24",
    number = "4",
    pages = "1077--1136",
    year = "2023"
}

@incollection {MR1941627,
    AUTHOR = {Joyce, Dominic},
     TITLE = {On counting special {L}agrangian homology 3-spheres},
 BOOKTITLE = {Topology and geometry: commemorating {SISTAG}},
    SERIES = {Contemp. Math.},
    VOLUME = {314},
     PAGES = {125--151},
 PUBLISHER = {Amer. Math. Soc.},
      YEAR = {2002},
      ISBN = {0-8218-2820-7},
   MRCLASS = {53C38 (32Q25 53C80 53D12 53D45)},
  MRNUMBER = {1941627},
MRREVIEWER = {Justin\ Sawon},
eprint={hep-th/9907013},
      archivePrefix={arXiv},
      primaryClass={hep-th},
       DOI = {10.1090/conm/314/05427},
       URL = {https://doi.org/10.1090/conm/314/05427},
}

@incollection {MR2222528,
    AUTHOR = {Katz, Sheldon},
     TITLE = {Gromov-{W}itten, {G}opakumar-{V}afa, and {D}onaldson-{T}homas
              invariants of {C}alabi-{Y}au threefolds},
 BOOKTITLE = {Snowbird lectures on string geometry},
    SERIES = {Contemp. Math.},
    VOLUME = {401},
     PAGES = {43--52},
 PUBLISHER = {Amer. Math. Soc.},
      YEAR = {2006},
      ISBN = {0-8218-3663-3},
   MRCLASS = {14N35 (14J32)},
  MRNUMBER = {2222528},
MRREVIEWER = {Hsian-Hua\ Tseng},
eprint={math/0408266},
      archivePrefix={arXiv},
      primaryClass={math.AG},
       DOI = {10.1090/conm/401/07552},
       URL = {https://doi.org/10.1090/conm/401/07552},
}

@article{Ooguri:2010yk,
    author = "Ooguri, Hirosi and Su\l{}kowski, Piotr and Yamazaki, Masahito",
    title = "{Wall Crossing As Seen By Matrix Models}",
    eprint = "1005.1293",
    archivePrefix = "arXiv",
    primaryClass = "hep-th",
    reportNumber = "AEI-2010-091, CALT-68-2786, IPMU10-0078",
    doi = "10.1007/s00220-011-1330-x",
    journal = "Commun. Math. Phys.",
    volume = "307",
    pages = "429--462",
    year = "2011"
}

@article{Mozgovoy:2020has,
    author = "Mozgovoy, Sergey and Pioline, Boris",
    title = "{Attractor invariants, brane tilings and crystals}",
    eprint = "2012.14358",
    archivePrefix = "arXiv",
    primaryClass = "hep-th",
    doi = "10.5802/aif.3682",
    journal = "Annales Inst. Fourier",
    volume = "75",
    number = "3",
    pages = "1331--1414",
    year = "2025"
}

@article{Jafferis:2008uf,
    author = "Jafferis, Daniel L. and Moore, Gregory W.",
    title = "{Wall crossing in local Calabi Yau manifolds}",
    eprint = "0810.4909",
    archivePrefix = "arXiv",
    primaryClass = "hep-th",
    reportNumber = "RUNHETC-2008-22",
    month = "10",
    year = "2008"
}

@article{Maulik:2003rzb,
    author = "Maulik, D. and Nekrasov, N. and Okounkov, A. and Pandharipande, R.",
    title = "{Gromov{\textendash}Witten theory and Donaldson{\textendash}Thomas theory, I}",
    eprint = "math/0312059",
    archivePrefix = "arXiv",
    reportNumber = "ITEP-TH-61-03, IHES-M-03-67",
    doi = "10.1112/S0010437X06002302",
    journal = "Compos. Math.",
    volume = "142",
    number = "05",
    pages = "1263--1285",
    year = "2006"
}

@article{Maulik:2004txy,
    author = "Maulik, D. and Nekrasov, N. and Okounkov, A. and Pandharipande, R.",
    title = "{Gromov{\textendash}Witten theory and Donaldson{\textendash}Thomas theory, II}",
    eprint = "math/0406092",
    archivePrefix = "arXiv",
    doi = "10.1112/S0010437X06002314",
    journal = "Compos. Math.",
    volume = "142",
    number = "05",
    pages = "1286--1304",
    year = "2006"
}

@article{Alim:2021mhp,
    author = "Alim, Murad and Saha, Arpan and Teschner, Joerg and Tulli, Iv{\'a}n",
    title = "{Mathematical Structures of Non-perturbative Topological String Theory: From GW to DT Invariants}",
    eprint = "2109.06878",
    archivePrefix = "arXiv",
    primaryClass = "hep-th",
    doi = "10.1007/s00220-022-04571-y",
    journal = "Commun. Math. Phys.",
    volume = "399",
    number = "2",
    pages = "1039--1101",
    year = "2023"
}

@article{Grassi:2022zuk,
    author = "Grassi, Alba and Hao, Qianyu and Neitzke, Andrew",
    title = "{Exponential Networks, WKB and Topological String}",
    eprint = "2201.11594",
    archivePrefix = "arXiv",
    primaryClass = "hep-th",
    reportNumber = "UTTG 31-2022, CERN-TH-2022-003",
    doi = "10.3842/SIGMA.2023.064",
    journal = "SIGMA",
    volume = "19",
    pages = "064",
    year = "2023"
}

@article{Szendr_i_2008,
   title="{Non-commutative Donaldson–Thomas invariants and the conifold}",
   volume={12},
   ISSN={1465-3060},
   url={http://dx.doi.org/10.2140/gt.2008.12.1171},
   DOI={10.2140/gt.2008.12.1171},
   number={2},
   journal={Geometry \& Topology},
   publisher={Mathematical Sciences Publishers},
   author={Szendrői, Balázs},
   year={2008},
   month=may, pages={1171–1202} 
}

@article{Fang:2016svw,
    author = "Fang, Bohan and Liu, Chiu-Chu Melissa and Zong, Zhengyu",
    title = "{On the Remodeling Conjecture for Toric Calabi-Yau 3-Orbifolds}",
    eprint = "1604.07123",
    archivePrefix = "arXiv",
    primaryClass = "math.AG",
    doi = "10.1090/jams/934",
    journal = "J. Am. Math. Soc.",
    volume = "33",
    number = "1",
    pages = "135--222",
    year = "2020"
}

@article{Fang:2013dna,
    author = "Fang, Bohan and Liu, Chiu-Chu Melissa and Zong, Zhengyu",
    editor = "Bouchard, Vincent and M{\'e}ndez-Diez, Stefan and Quigley, Callum and Doran, Charles",
    title = "{All Genus Open-Closed Mirror Symmetry for Affine Toric Calabi-Yau 3-Orbifolds}",
    eprint = "1310.4818",
    archivePrefix = "arXiv",
    primaryClass = "math.AG",
    journal = "Proc. Symp. Pure Math.",
    volume = "93",
    pages = "1",
    year = "2015"
}

@article{Nekrasov:2003rj,
    author = "Nekrasov, Nikita and Okounkov, Andrei",
    title = "{Seiberg-Witten theory and random partitions}",
    eprint = "hep-th/0306238",
    archivePrefix = "arXiv",
    reportNumber = "ITEP-TH-36-03, PUDM-2003, IHES-P-03-43",
    doi = "10.1007/0-8176-4467-9_15",
    journal = "Prog. Math.",
    volume = "244",
    pages = "525--596",
    year = "2006"
}

@article{Iwaki:2020efz,
    author = "Iwaki, Kohei and Kidwai, Omar",
    title = "{Topological recursion and uncoupled BPS structures I: BPS spectrum and free energies}",
    eprint = "2010.05596",
    archivePrefix = "arXiv",
    primaryClass = "math-ph",
    doi = "10.1016/j.aim.2022.108191",
    journal = "Adv. Math.",
    volume = "398",
    pages = "108191",
    year = "2022"
}

@article{Iwaki:2021zif,
    author = "Iwaki, Kohei and Kidwai, Omar",
    title = "{Topological Recursion and Uncoupled BPS Structures II: Voros Symbols and the $\tau $-Function}",
    eprint = "2108.06995",
    archivePrefix = "arXiv",
    primaryClass = "math-ph",
    doi = "10.1007/s00220-022-04563-y",
    journal = "Commun. Math. Phys.",
    volume = "399",
    number = "1",
    pages = "519--572",
    year = "2023"
}

@article {MR3713014,
    AUTHOR = {Cazzaniga, Alberto and Morrison, Andrew and Pym, Brent and
              Szendr\"oi, Bal\'azs},
     TITLE = "{Motivic {D}onaldson-{T}homas invariants of some quantized
              threefolds}",
   JOURNAL = {J. Noncommut. Geom.},
  FJOURNAL = {Journal of Noncommutative Geometry},
    VOLUME = {11},
      YEAR = {2017},
    NUMBER = {3},
     PAGES = {1115--1139},
      ISSN = {1661-6952,1661-6960},
   MRCLASS = {14N35 (14A22 14J32 16G20)},
  MRNUMBER = {3713014},
MRREVIEWER = {Amin\ Gholampour},
eprint={1510.08116},
      archivePrefix={arXiv},
      primaryClass={math.AG},
       DOI = {10.4171/JNCG/11-3-10},
       URL = {https://doi.org/10.4171/JNCG/11-3-10},
}

@article {MR3000467,
    AUTHOR = {Morrison, Andrew},
     TITLE = {Motivic invariants of quivers via dimensional reduction},
   JOURNAL = {Selecta Math.},
  FJOURNAL = {Selecta Mathematica. New Series},
    VOLUME = {18},
      YEAR = {2012},
    NUMBER = {4},
     PAGES = {779--797},
      ISSN = {1022-1824,1420-9020},
   MRCLASS = {16G20 (14N35)},
  MRNUMBER = {3000467},
MRREVIEWER = {M\'aty\'as\ Domokos},
 eprint={1103.3819},
      archivePrefix={arXiv},
      primaryClass={math.AG},
       DOI = {10.1007/s00029-011-0081-z},
       URL = {https://doi.org/10.1007/s00029-011-0081-z},
}

@article {MR2951762,
    AUTHOR = {Joyce, Dominic and Song, Yinan},
     TITLE = {A theory of generalized {D}onaldson-{T}homas invariants},
   JOURNAL = {Mem. Amer. Math. Soc.},
  FJOURNAL = {Memoirs of the American Mathematical Society},
    VOLUME = {217},
      YEAR = {2012},
    NUMBER = {1020},
     PAGES = {iv+199},
      ISSN = {0065-9266,1947-6221},
      ISBN = {978-0-8218-5279-8},
   MRCLASS = {14N35 (14D23 14F05 14J32)},
  MRNUMBER = {2951762},
MRREVIEWER = {Amin\ Gholampour},
eprint={0810.5645},
      archivePrefix={arXiv},
      primaryClass={math.AG},
       DOI = {10.1090/S0065-9266-2011-00630-1},
       URL = {https://doi.org/10.1090/S0065-9266-2011-00630-1},
}

@article {MR2600874,
    AUTHOR = {Behrend, Kai},
     TITLE = "{Donaldson-{T}homas type invariants via microlocal geometry}",
   JOURNAL = {Ann. of Math. (2)},
  FJOURNAL = {Annals of Mathematics. Second Series},
    VOLUME = {170},
      YEAR = {2009},
    NUMBER = {3},
     PAGES = {1307--1338},
      ISSN = {0003-486X,1939-8980},
   MRCLASS = {14N35 (14C15 14C17 14D23)},
  MRNUMBER = {2600874},
MRREVIEWER = {Hsian-Hua\ Tseng},
eprint={math/0507523},
      archivePrefix={arXiv},
      primaryClass={math.AG},
       DOI = {10.4007/annals.2009.170.1307},
       URL = {https://doi.org/10.4007/annals.2009.170.1307},
}

@article{Nekrasov:2016qym,
    author = "Nekrasov, Nikita",
    title = "{BPS/CFT correspondence II: instantons at crossroads, moduli and compactness theorem}",
    eprint = "1608.07272",
    archivePrefix = "arXiv",
    primaryClass = "hep-th",
    doi = "10.4310/ATMP.2017.v21.n2.a4",
    journal = "Adv. Theor. Math. Phys.",
    volume = "21",
    pages = "503--583",
    year = "2017"
}

@article{Eager:2016yxd,
    author = "Eager, Richard and Selmani, Sam Alexandre and Walcher, Johannes",
    title = "{Exponential Networks and Representations of Quivers}",
    eprint = "1611.06177",
    archivePrefix = "arXiv",
    primaryClass = "hep-th",
    doi = "10.1007/JHEP08(2017)063",
    journal = "JHEP",
    volume = "08",
    pages = "063",
    year = "2017"
}

@article{Banerjee:2024smk,
    author = "Banerjee, Sibasish and Romo, Mauricio and Senghaas, Raphael and Walcher, Johannes",
    title = "{Exponential networks for linear partitions}",
    eprint = "2403.14588",
    archivePrefix = "arXiv",
    primaryClass = "hep-th",
    doi = "10.21468/SciPostPhys.18.4.128",
    journal = "SciPost Phys.",
    volume = "18",
    number = "4",
    pages = "128",
    year = "2025"
}

@article{kontsevich2008stabilitystructuresmotivicdonaldsonthomas,
      title="{Stability structures, motivic Donaldson-Thomas invariants and cluster transformations}", 
      author={Maxim Kontsevich and Yan Soibelman},
      year={2008},
      eprint={0811.2435},
      archivePrefix={arXiv},
      primaryClass={math.AG},
}

@article{Beaujard:2020sgs,
    author = "Beaujard, Guillaume and Manschot, Jan and Pioline, Boris",
    title = "{Vafa{\textendash}Witten Invariants from Exceptional Collections}",
    eprint = "2004.14466",
    archivePrefix = "arXiv",
    primaryClass = "hep-th",
    doi = "10.1007/s00220-021-04074-2",
    journal = "Commun. Math. Phys.",
    volume = "385",
    number = "1",
    pages = "101--226",
    year = "2021"
}

@article{Awata:2008ed,
    author = "Awata, Hidetoshi and Kanno, Hiroaki",
    title = "{Refined BPS state counting from Nekrasov's formula and Macdonald functions}",
    eprint = "0805.0191",
    archivePrefix = "arXiv",
    primaryClass = "hep-th",
    doi = "10.1142/S0217751X09043006",
    journal = "Int. J. Mod. Phys. A",
    volume = "24",
    pages = "2253--2306",
    year = "2009"
}

@article{Sulkowski:2009rw,
    author = "Sulkowski, Piotr",
    title = "{Wall-crossing, free fermions and crystal melting}",
    eprint = "0910.5485",
    archivePrefix = "arXiv",
    primaryClass = "hep-th",
    reportNumber = "CALT-68-2756",
    doi = "10.1007/s00220-010-1153-1",
    journal = "Commun. Math. Phys.",
    volume = "301",
    pages = "517--562",
    year = "2011"
}

@article{Nekrasov:2003af,
    author = "Nekrasov, Nikita A.",
    title = "{Seiberg-Witten prepotential from instanton counting}",
    booktitle = "{International Congress of Mathematicians}",
    eprint = "hep-th/0306211",
    archivePrefix = "arXiv",
    month = "6",
    year = "2003"
}

@article{Nekrasov:2004ws,
    author = "Nekrasov, N. A.",
    editor = "Fukuma, M. and Itoyama, H. and Nakatsu, T. and Tsuchiya, A.",
    title = "{Solution of N=2 gauge theory}",
    doi = "10.1143/PTPS.152.73",
    journal = "Prog. Theor. Phys. Suppl.",
    volume = "152",
    pages = "73--79",
    year = "2004"
}

@article{Eguchi:2003sj,
    author = "Eguchi, Tohru and Kanno, Hiroaki",
    title = "{Topological strings and Nekrasov's formulas}",
    eprint = "hep-th/0310235",
    archivePrefix = "arXiv",
    reportNumber = "UT-03-35",
    doi = "10.1088/1126-6708/2003/12/006",
    journal = "JHEP",
    volume = "12",
    pages = "006",
    year = "2003"
}
\end{document}